\documentclass[letterpaper,11pt,twoside,keywordsasfootnote,addressatend,noinfoline]{article}
\usepackage{fullpage}
\usepackage[english]{babel}
\usepackage{amssymb}
\usepackage{amsmath}
\usepackage{theorem}
\usepackage{epsfig}
\usepackage{enumitem}
\usepackage{mathrsfs}
\usepackage{rotating}
\usepackage{imsart}
\usepackage{color}

\theorembodyfont{\slshape}
\newtheorem{theorem}{Theorem}

\newtheorem{lemma}[theorem]{Lemma}

\newenvironment{proof}{\noindent{\scshape Proof.}}{\hspace*{2mm} $\square$}

\newcommand{\C}{\mathscr{C}}
\newcommand{\N}{\mathbb{N}}

\newcommand{\T}{\mathbb{T}}
\renewcommand{\S}{\mathbb{S}}
\newcommand{\Or}{\mathcal{O}}
\newcommand{\ind}{\mathbf{1}}

\newcommand{\n}{\hspace*{-6pt}}

\DeclareMathOperator{\card}{card \,}
\DeclareMathOperator{\bernoulli}{Bernoulli \,}
\DeclareMathOperator{\uniform}{Uniform \,}

\DeclareMathOperator{\var}{Var}
\DeclareMathOperator{\cov}{cov}

\usepackage{changepage}

\usepackage{hyperref}

\usepackage{multirow}
\usepackage{arydshln}

\begin{document}

\begin{frontmatter}
\title     {Probabilistic Framework For Loss Distribution \\ Of Smart Contract Risk\thanks{{The
            numerical results presented in this work are produced by the joint invention of the authors.
            The invention is patent pending under the heading
            ``Systems and methods for a simulation program of percolation model for the loss distribution of smart contracts caused by a cyber attack or contagious failure''.}}}
\runtitle  {Percolation framework for the loss distribution of smart contract risks}
\author    {\large Petar Jevti\'c and Nicolas Lanchier}
\runauthor {Petar Jevti\'c and Nicolas Lanchier}
\address   {School of Mathematical and Statistical Sciences \\ Arizona State University \\ Tempe, AZ 85287, USA. \\ petar.jevtic@asu.edu \\ nicolas.lanchier@asu.edu} \bigskip

\maketitle

\begin{abstract} \
{Smart contract risk can be defined as a financial risk of loss due to cyber attacks on or contagious failures of smart contracts. Its quantification is of paramount importance to technology platform providers as well as companies and individuals when considering the deployment of this new technology. That is why, as our primary contribution, we propose a structural framework of aggregate loss distribution for
 smart contract risk under the assumption of a tree-stars graph topology representing the network of interactions among smart contracts and their users.}
 Up to our knowledge, there exist no theoretical frameworks or models of an aggregate loss distribution for smart contracts in this setting.
 To achieve our goal, we contextualize the problem in the probabilistic graph-theoretical framework using bond percolation models.
 We assume that the smart contract network topology is represented by a random tree graph of finite size, and that each smart contract
 is the center of a {random} star graph whose leaves represent the users of the smart contract.
 We allow for heterogeneous loss topology superimposed on this smart contract and user topology and provide analytical results and instructive
 numerical examples.
\end{abstract}



\end{frontmatter}

%\kwd{Smart contracts, cyber risk, operational risk, loss modeling, percolation model, random graphs, contagion, network topology, smart contract insurance.}

%%%%%%%%%%%%%%%%%%%%%%%%%%%%%%%%%%%%%%%%%%%%%%%%%%%%%%%%%%%%%%%%%%%%%%%%%%%%%%%%%%%%%%%%%%%%%%%%%%%%%%%%%%%%%%%%%%%%%%%%%%%%%%%%%%%%%%%%%%%%%%%%%%%%%%%%%%%%%%%%%%%%%%%%%%%%

\section{Introduction}

\noindent{\bf Technology.} In its core, blockchain technology represents an open but distributed ledger where transactions between parties are recorded in verifiable and
 immutable ways~\cite{iansiti2017truth}.
 Blockchains emerged in global public spheres in 2008 with the advent of Bitcoin digital currency, and was conceptually and analytically founded
 by the now legendary work of~\cite{nakamoto2008bitcoin}.
 
\indent Prescient when voiced and skillfully phrased by~\cite{szabo1996smart}, smart contracts are first defined as~``a set of promises, specified in
 digital form, including protocols within which the parties perform on these promises''.
 From a software engineering perspective, smart contracts can be described as self-executing scripts running on blockchain platforms that can be
 both private, public or consortium and semi-private blockchains. \smallskip
 
\noindent{\bf Present and future.} Today, the penetration of blockchain technology is a wide spread phenomenon across industries and its use is accelerating~\cite{insights2016banking}. The largest public blockchain platform that offers smart contract capabilities is Ethereum.
{The digital currency of the Etherium platform is Ether (ETH) which has market capitalization\footnote{See \nolinkurl{https://coinmarketcap.com/currencies/ethereum/}.}
 around 27 billion~USD, with 24h volume of trade around 9 billion~USD.}
 The number\footnote{See \nolinkurl{https://cointelegraph.com/news/ethereum-smart-contracts-up-75-to-almost-2m-in-march}.} of platform hosted smart
 contracts, i.e., blockchain stored scripts that can be coded in Solidity language, has recently reached almost 2 million. 

The promises of increased efficiencies of economic transactions and automated interactions between economical agents, novel ways of
 resource utilization and monetization, data integrity and privacy~\cite{shrier2016blockchain}, etc. are tantalizing (see~\cite{swan2015blockchain}
 or~\cite{van2018blockchain}). Blockchain enabled technologies, which includes smart contract technology, are estimated to produce business value-add growth by~2025 ranging
 in~176 billion~USD\footnote{See Lovelock, J. and  Furlonger, D., 2017. Three Things CIOs Need to Know About the Blockchain Business Value Forecast.
 Published by Gartner.}. Most strikingly, the World Economic Forum  \cite{shift2015technology} survey of 800 information and communications
 executives and experts reveals belief that around ten percent of global GDP would be found on blockchain systems by year 2027. Therefore, it is increasingly being recognized that, associated with digital assets, in conjunction with smart contracts, blockchain technology offers
 novel ways of organizing the economy and even the society across myriad of everyday interactions. \smallskip
 %In insurance industry, the ramifications of this technology have been recently discussed in~\cite{gatteschi2018blockchain, hans2017blockchain}.

\noindent {\bf Risk.}  As any novel technology, the smart contract technology comes with its own risks~\cite{nikolic2018finding} that expose its users to
 potentially unforeseen liabilities.
 Contagious losses can originate from many sources starting with coding errors~\cite{atzei2017survey}, malicious cyber attacks~\cite{marcus2018low}
 or even under-optimized smart contracts~\cite{chen2017under}.
Notwithstanding the smaller ones, the losses can be of considerable size, and the now infamous~2016 Ethereum DAO attack~\cite{mehar2019understanding}, where over~50 million~USD worth of
 Ethereum were misappropriated, looms large as an example of a potential liability.
 Another example is the 2017 parity multi-signature wallet attack\footnote{See \nolinkurl{https://cointelegraph.com/news/parity-multisig-wallet-hacked-or-how-come}}
 where around 30 million~USD then equivalent value of Ethers was stolen and subsequently due to exploited code vulnerability the equivalent of around 150
 million~USD permanently rendered inaccessible.
 Also, in 2018, MyEtherWallet\footnote{See \nolinkurl{https://www.theverge.com/2018/4/24/17275982/myetherwallet-hack-bgp-dns-hijacking-stolen-ethereum}.}
 had its about 17 million~USD worth in Ether stolen.
Sadly, the future does not appear without clouds as the current research suggests.
 The recent findings \cite{nikolic2018finding} show that at least around 30,000 current Ethereum smart contracts are at risk due to their
 particular characteristics. 
 
That is why in this work, we define \textit{a smart contract risk} as \textit{a risk of financial loss due to cyber attacks or contagious failures of smart contracts}.
 The risk can originate from the smart contract under consideration, or its users, or other smart contracts the smart contract under consideration communicates with
 during the course of its execution, or their users.
 The losses may be the result of misappropriation or misallocation of funds belonging to wallets of users or smart contracts under consideration. Consequently, from management perspective, if liability is left poorly understood, the risks arising from application of this novel technology may jeopardize platform providers or stifle decisions for their faster adoption.\smallskip

\noindent  {{\bf Challenge.} The characterization of loss distribution is widely used approach for quantification of the frequency and severity distributions of operational risk losses (see \cite{shevchenko2011modelling}). } In practice, the empirical loss distribution becomes available after sufficient time has passed so that sufficiently large number of loss observations can be collected. Unfortunately, in the case of smart contract risk, due to the lack of data, there exist no empirical loss distributions in proper sense .
 What is currently available is a handful of recorded losses spread across smart contract platforms and partially recorded as anecdotes in the academic literature.
 In particular, there is insufficient information for the creation of empirical loss distributions and thus characterization of the risk from
 a statistical/empirical perspective.
 % and their use in conjunction with risk measure to price coverages on smart contracts.
 Sadly, in short term, the future offers no hope here.
 In fact, this situation suggests a different approach, namely the creation of structural models for the loss distribution, which this work addresses.
 To our knowledge, this is the first work  that is concerned with the characterization of smart contract risks from probabilistic perspective 
and develops credible and practical structural models for the loss distribution. As such, this work paves the way for insurers to price smart contract risks, which is highly relevant in decisions for creating new smart contract risk related insurance product lines. \smallskip %\vspace*{5pt} \\

\noindent {\bf Mathematical Conceptualization.} Conceptually, we envision the smart contract under consideration as the root vertex of a random tree call graph~\cite{frowis2017code,ryder1979constructing}.
 Call graphs are comprised of vertices that are smart contracts interacting directly or indirectly with the smart contract under consideration during its execution.
 These smart contracts can be seen as offspring of the root smart contract in an undirected tree graph.
 For their proper performance, these smart contracts might rely on the execution of some other smart contracts which they call.
 Those would be, in turn, their offspring and so on up to some distance from the root smart contract.
 Here, random graphs are used to conceptualize the dynamical nature of call graphs.
 At any given time, the smart contract under consideration can have different patterns of communication with some (or none) offspring smart contracts
 which, in turn, might have smart contracts they communicate with in temporally inhomogeneous ways.
 The authors of~\cite{frowis2017code} investigated nearly~200,000 smart contracts on Etherium platform and, among those that call other smart
 contracts, which was a majority, they found that only a small number of call graphs had loops.
%  Thus, the need for tree graph form of smart contracts emerges.
 This motivates the use of tree graphs, i.e., graphs with no loop, to model the network of smart contracts.
 Each of the smart contracts in this structure may in principle have users it interacts with.
 Assuming that the users are not shared among different smart contracts, the random tree-stars graph structure naturally emerges.  
 
% Empirically, the rationale for tree graph structure of smart contracts was found in the work of ... who
\indent We use a two-parameter bond percolation model to describe the contagion process among smart contracts and their users.
 Bond percolation was introduced in~\cite{broadbent_hammersley_1957}.
 For a pedagogical and thorough introduction to percolation, we refer the reader to~\cite{grimmett_1989}, while a brief overview of the
 main results is available in~\cite[chapter~13]{lanchier2017stochastic}.
% The choice of bond percolation as a model of contagion is relevant for actuaries and consistent with actuarial modeling practices.
 There is a wide variety of contagion processes one can choose from as a model component in framework building.
However, to our knowledge, there is no extensive empirical study of contagions in smart contracts~(certainly a valuable research question to be
 addressed in the future) so the choice of contagion process is left to the modelers.
That is why, given the lack of an empirical study, it is natural to assume that the way smart contracts interact is strongly influenced by
 random factors and the topology of the call graph.
 All considered, in our framework we use bond percolation as a starting model of contagion among the models that account for both stochasticity and network structure, and leave other choices to future research. 
 
%  {All considered, in our framework we use bond percolation as a starting model of contagion among the models that account for both stochasticity and network structure, and leave other choices to future research.} \\
%  We use bond percolations as one of the simples contagion models that allows for analytical tractability. \\
% \indent As a final modeling choice, we also assume that an arrangement of monetary assets exists and that it follows the topology of the network, i.e.,
%  we attach a monetary asset with a certain value to each node of the network, either of user or of smart contract type.

\indent As a final modeling choice in our proposed framework, we also include a configuration of monetary assets on the network, i.e., we attach a monetary asset with a certain
dynamic value to each node of the network, either of user type or of smart contract type.
 Simply put, we assume that users and smart contracts, in their wallets, hold some assets that have a monetary value at any given time.
 This arrangement of monetary values across the network constitutes a {cost topology}.
 The compromise of a node in the network {(due to a cyber attack, an operational failure, etc.)} entails the loss of the monetary asset and its value.
 To account for the dynamical nature of these assets across time and over the evolving network, we assume that the asset values are
 represented by random variables.
 The percolation model then defines the contagion process stemming from {the event of a node being compromised} given a particular temporal instance
 of the tree-stars network topology.
 Finally, the sum of all the losses, given the particular first node being compromised and the realization of the associated contagion process,
 characterizes one observation point in the aggregate loss distribution due to cyber attacks or operational failures of smart contracts. 
 
\indent In this setting, we give analytical and numerical results related to the mean and the variance of the aggregate loss distribution.
 We emphasize that our results hold for arbitrarily large random tree-stars graphs, and for all possible choices of the parameters of the bond percolation
 model and the distribution of the asset values. \\
 
\noindent The rest of this paper is organized as follows.
 In Section~\ref{sec:model}, the mathematical framework to model loss due to cyber attacks and/or {operational} failures is developed.
 Section~\ref{sec:results} presents the main analytical and numerical results about the mean and variance of the loss distribution.
%  The numerical results are given in Section~\ref{sec:numerics} while Section~\ref{sec:conclusion} concludes the paper.
 The remaining section contains the proofs of the analytical results.

%broadbent_hammersley_1957
%grimmett_1989
%lanchier2017stochastic

%\cite{atzei2018sok}

%%%%%%%%%%%%%%%%%%%%%%%%%%%%%%%%%%%%%%%%%%%%%%%%%%%%%%%%%%%%%%%%%%%%%%%%%%%%%%%%%%%%%%%%%%%%%%%%%%%%%%%%%%%%%%%%%%%%%%%%%%%%%%%%%%%%%%%%%%%%%%%%%%%%%%%%%%%%%%%%%%%%%%%%%%%%

\section{Framework for the aggregate loss}\label{sec:model}

\begin{figure}[h!]
\includegraphics[width=350pt]{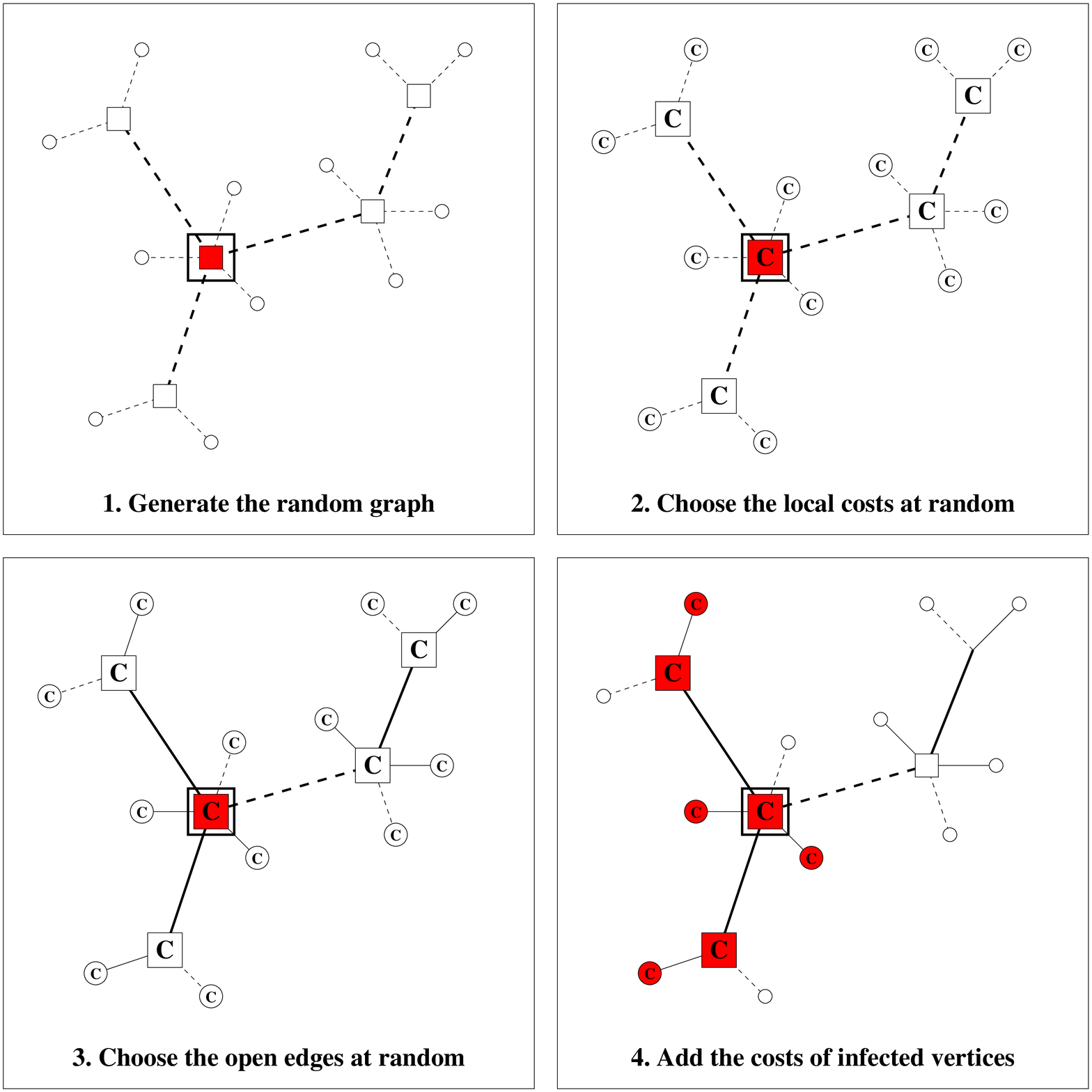}
\caption{\upshape{Illustration of the random process generating a single contagion and the associated costs.
                  The squares represent the smart contracts while the circles represent the users of the smart contracts.
                  In each picture, the square with a frame is the smart contract under consideration whose loss distribution is considered.
                  The pictures are for scenario~1 but the process is the same for the other scenarios except to compute the cost.
                  First, we generate the random tree of smart contracts and add the random stars of users.
                  Second, we choose local costs at random for all the vertices.
                  Third, we use independent coin flips to determine the open edges.
                  Fourth, we add the costs of all the vertices that are {compromised} (connected to the origin by a path of open
                  edges) and circled in dashed lines in Figure~\ref{fig:scenarios}.}}
\label{fig:dynamics}
\end{figure}

We model the aggregate loss up to time~$t$
 using a continuous-time Markov chain~$(L_t)$ that consists of the combination of a Poisson process representing the times at which contagions strike,
{the} random graph representing the evolving connections among smart contracts and users, a percolation process on this random graph modeling the
 spread of the contagion, and a collection of independent random variables on the vertex set representing the evolving monetary assets.
 For the purpose of understanding the main characteristics of loss distribution and risk pricing, the main objective is to study the mean and the variance of the
 random variable~$L_t$. \\
\indent From a probabilistic perspective, random graphs relevant for our problem consist of the composition of a random tree and a
 collection of random stars.
 The former models the connections among the smart contracts whereas the latter models the connections between smart contracts and users.
 Because the network consists of a two layers, it is natural to include two percolation parameters:
 one parameter representing the probability that the contagion spreads across an edge connecting two smart contracts and another parameter representing
 the probability that the contagion spreads across an edge connecting a smart contract and a user.
 Similarly, we consider two different distributions for the local costs~(monetary assets), one modeling the loss resulting from a smart contract's wallet
 being compromised, and another one modeling the loss resulting from a user's wallet being compromised. More precisely, the process is constructed using the following components:
% \begin{figure}[t]
% \includegraphics[width=340pt]{tree-star.eps}
% \caption{\upshape{Example of realization of the graph~$G = (V, E)$ with radius~$R = 2$.
%          Black and white vertices represent the smart contracts and the users, respectively.
%          The thick solid/dashed lines are the edges in~$E_+$ connecting smart contracts, while the thin solid/dashed lines are the edges in~$E_-$.
%          Along with a realization of the random graph, our picture shows a possible realization of the bond percolation process on this graph where
%          the solid lines are the open edges and the dashed lines the closed edges.
%          The vertices that are circled are the ones in the open cluster starting at the root, and so the ones that would be infected in the event of
%          a contagion starting at the root.}}
% \label{fig:tree-star}
% \end{figure}
\begin{itemize}
 \item A Poisson process~$(N_t)$ with intensity~$\lambda$. \vspace*{2pt}
 \item A random graph~$G = (V, E)$ consisting of the combination of a random rooted tree with radius~$R$ and offspring distribution described by a random
       variable~$X_+$, and random stars with degree described by a random variable~$X_-$, with probability mass functions
       $$ P (X_+ = k) = p_k \quad \hbox{and} \quad P (X_- = k) = q_k \quad \hbox{for all} \quad k \in \N. $$
 \item Two percolation parameters~$p, q \in (0, 1)$. \vspace*{2pt}
 \item A random variable~$\widehat C_+$ describing the loss due to a smart contract being {compromised}. \vspace*{2pt}
 \item A random variable~$\widehat C_-$ describing the loss due to a user being {compromised}.
\end{itemize}
 The process evolves as follows.
 At the arrival times
 $$ T^i = \inf \{t : N_t = i \}, \quad i > 0, $$
 of the Poisson process, we let~$G^i = (V^i, E^i)$ be a realization of the random graph modeling the connections among smart contracts and users
 at the time~$T^i$ of the~$i$th contagion. \\
\indent To construct this random graph, we draw~$X_+$ edges starting from a root~0, meaning~$k$ edges with probability~$p_k$, and additional edges starting
 from each of the subsequent vertices using the same probability distribution.
 The construction stops after~$R$ steps, which results in
 $$ \T_R^i = (V_+^i, E_+^i) = \hbox{random tree with radius at most~$R$}, $$
 where~$V_+^i$ represents the set of smart contracts.
 Then, from each smart contract~$x \in V_+^i$, we independently draw~$X_-$ edges, meaning~$k$ edges with probability~$q_k$, thus creating
 $$ \S^i (x) = (V_-^i (x), E_-^i (x)) = \hbox{random star with center~$x$} \quad \hbox{for all~$x \in V_+^i$}. $$
 The leaves of the star represent the users connected to smart contract~$x$, and we assume that each user is connected to only one
 smart contract.
 Letting~$V_-^i$ be the set of all users, and~$E_-^i$ be the set of all edges connecting a user to a smart contract, the construction
 results in a random graph
 $$ G^i = (V^i, E^i) \quad \hbox{where} \quad V^i = V_+^i \cup V_-^i \ \hbox{and} \ E^i = E_+^i \cup E_-^i. $$
 See the top-left panel of Figure~\ref{fig:dynamics} for a picture where the squares represent the smart contracts and the circles represent the users. \\
\indent To quantify the financial loss, we attach a random local cost~$\widehat C_y^i$ to each vertex~$y \in V^i$ representing the
 loss resulting from vertex~$y$ being {compromised}.
 More precisely, we let
 $$ \begin{array}{rclcl}
    \widehat C_y^i & \n = \n & \widehat C_+ & \hbox{in distribution} & \hbox{for all} \ y \in V_+^i \vspace*{4pt} \\
    \widehat C_y^i & \n = \n & \widehat C_- & \hbox{in distribution} & \hbox{for all} \ y \in V_-^i \end{array} $$
 be independent.
 Considering two different distributions for the local costs is motivated by the fact that the loss due to a smart contract being {compromised} in principle may be significantly
 different from the loss due to a user being {compromised}.
 See the top-right panel of Figure~\ref{fig:dynamics} for a picture. \\
\indent To model the contagion itself, we use the framework of percolation theory, and more precisely, bond percolation (percolation on the edges).
 That is, we let
 $$ \begin{array}{rcl}
    \xi^i (e) = \bernoulli (p) & \hbox{for all} & e \in E_+^i \vspace*{4pt} \\
    \xi^i (e) = \bernoulli (q) & \hbox{for all} & e \in E_-^i \end{array} $$
 be independent.
 Following the terminology of percolation theory, edges with~$\xi^i (e) = 1$ are said to be open.
 See the bottom-left panel of Figure~\ref{fig:dynamics} for a picture where the solid edges are open and the dashed edges are closed.
 Given that the contagion starts at vertex~$\Or^i$, which we call from now on the origin of the contagion, the set of vertices that get {compromised} is
 $$ \C^i (\Or^i) = \{y \in V^i : \ \hbox{there is a path of open edges connecting~$\Or^i$ and~$y$} \}, $$
 called the open cluster starting at~$\Or^i$.
 See the bottom-right panel of Figure~\ref{fig:dynamics} for a picture where the open cluster starting at the root is represented in red. \\
\indent For the purpose of loss modeling, we are only interested in the vertices being compromised and their cost in certain subsets depending on the
 origin of the contagion.
 Therefore, instead of just considering the total size and the total cost of the contagion, we define more generally two collections of random
 variables as follows.
 For every subset~$A^i \subset V^i$, we let~$S^i (A^i)$ be the number of vertices in~$A^i$ that are {compromised} at time~$T^i$.
 In equation, this can be written as
 $$ S^i (A^i) = \card (\C^i (\Or^i) \cap A^i) \quad \hbox{for all} \quad A^i \subset V^i. $$
 Similarly, we define the financial loss restricted to subset~$A^i \subset V^i$ as
%  $$ C^i (A^i) = \sum_{y \in V^i} \,\widehat C_y^i \ \ind \{y \in \C^i (\Or^i) \cap A^i \} \quad \hbox{for all} \quad A^i \subset V^i, $$
 $$ C^i (A^i) = \sum_{y \in S^i (A^i)} \widehat C_y^i \quad \hbox{for all} \quad A^i \subset V^i, $$
 the sum of the local costs of all the vertices that are in subset~$A^i$ and that are {compromised}, i.e., in the open cluster starting at the origin~$\Or^i$ of
 the contagion. \\
\begin{figure}[t]
\includegraphics[width=350pt]{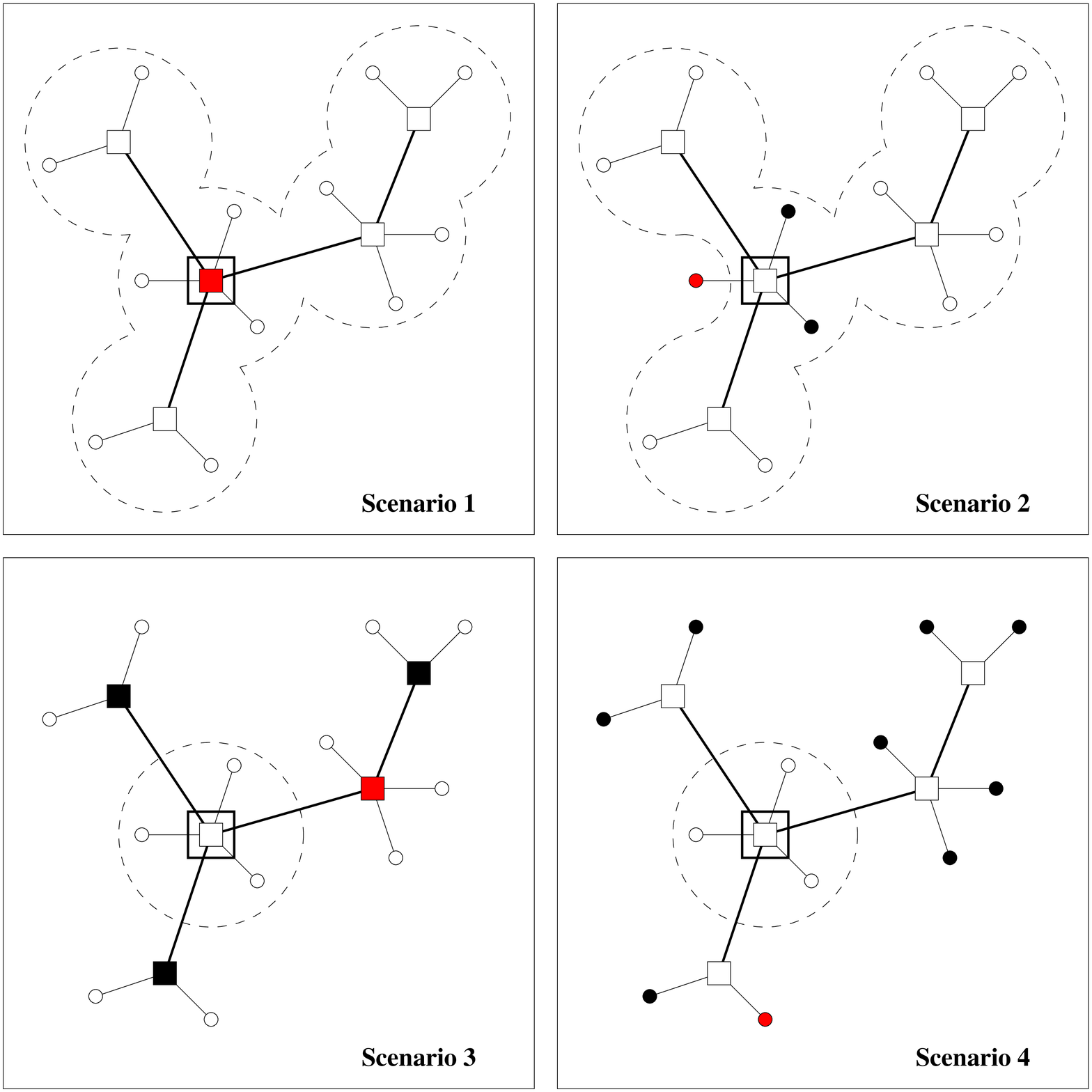}
\caption{\upshape{Pictures of the four scenarios.
                  In each picture, the red vertex indicates the possible origin of the contagion and the black vertices (if any) the other vertices from
                  which the contagion can start.
                  The four resulting subsets of red and black vertices form a partition of the vertex set:
                  only the smart contract at the root, only the users of the smart contract at the root, all the smart contracts except the
                  root, and all the users except the users of the root.
                  In each scenario, the relevant cost from the point of view of the smart contract at the root is the loss restricted to the subset
                  of vertices circled in dashed lines.}}
\label{fig:scenarios}
\end{figure}
\indent To complete the mathematical description of the loss resulting from a single contagion, we still need to explain how the origin~$\Or^i$ and the
 subset~$A^i$ are chosen.
 There are four distinct risk scenarios, and we assume that scenario~$j$ occurs with probability~$Q_j$ at each arrival time of the~Poisson process
 independently of everything else.
 The four scenarios are as follows.
\begin{enumerate}
 \item The contagion is due to the smart contract at the root being compromised.
       In this case, the origin of the contagion is the root and the loss is the total loss over all the network so
       $$ \Or^i = 0 \quad \hbox{and} \quad A^i = V^i. $$
 \item The contagion is due to a user of the smart contract at the root compromising this smart contract.
       In this case, the origin is chosen uniformly at random from the set of users of the root, and the loss that is of interest is the total loss except
       for the user which originates the compromising activity, so
       $$ \Or^i = \uniform (V_-^i (0) \setminus \{0 \}) \quad \hbox{and} \quad A^i = V^i \setminus \{\Or^i \}. $$
 \item The contagion is due to one of the smart contracts excluding the root being compromised.
       In this case, the origin of the contagion is chosen at random from the set of smart contracts other than the root and, from the perspective
       of the smart contract at the root, the loss which is of interest is the loss restricted to the smart contract at the root and its users so
       $$ \Or^i = \uniform (V_+^i \setminus \{0 \}) \quad \hbox{and} \quad A^i = V_-^i (0). $$
 \item The contagion is due to a user of one of the smart contracts other than the root compromising this smart contract.
       In this case, the origin is chosen uniformly at random from the set of users of the smart contracts other than the root, and the loss which is of
       interest is again the loss restricted to the smart contract at the root and its users so
       $$ \Or^i = \uniform (V_-^i \setminus V_-^i (0)) \quad \hbox{and} \quad A^i = V_-^i (0). $$
\end{enumerate}
 See Figure~\ref{fig:scenarios} for a picture of the four scenarios.
{Note that the four sets representing the possible origins of the contagion in the four scenarios form a partition of the network, therefore our model and
 analysis cover all possibilities.} \smallskip

\indent Finally, the random variable~$L_t$ is defined as the aggregate financial loss caused by all the contagions that occur between time zero and time~$t$.
 In equation,
%  $$ L_t = \sum_{i = 1}^{N_t} \,C^i (A^i) = \sum_{i = 1}^{N_t} \sum_{y \in V^i} \,\widehat C_y^i \ \ind \{y \in \C^i (\Or^i) \cap A^i \}. $$
 $$ L_t = \sum_{i = 1}^{N_t} \,C^i (A^i) = \sum_{i = 1}^{N_t} \sum_{y \in S^i (A^i)} \widehat C_y^i. $$
 For the purpose of loss distribution characterization and risk pricing, the main objective is to compute the expected value and the variance of the aggregate loss~$L_t$.
 Since the financial losses resulting from different contagions are independent, and because the losses resulting from contagions of the same type~$j$ are
 identically distributed, the mean and variance of the aggregate loss~$L_t$ can be  deduced from the mean and variance of the loss resulting
 from a single contagion.
 In particular, we first focus on the loss resulting from a single contagion of type~$j$, and drop all the superscripts~$i$ referring to the number of the
 contagion to avoid cumbersome notations.

%%%%%%%%%%%%%%%%%%%%%%%%%%%%%%%%%%%%%%%%%%%%%%%%%%%%%%%%%%%%%%%%%%%%%%%%%%%%%%%%%%%%%%%%%%%%%%%%%%%%%%%%%%%%%%%%%%%%%%%%%%%%%%%%%%%%%%%%%%%%%%%%%%%%%%%%%%%%%%%%%%%%%%%%%%%%

\section{Main results}
\label{sec:results}
 This section presents our analytical and numerical results about the loss distribution. \smallskip

\noindent {\bf Analytical results.}
 As previously mentioned, we first study the loss resulting from a single contagion in the context of scenario~$j$, for~$1 \leq j \leq 4$.
 To keep the notation short, we write
 $$ S (V) = S, \quad S (V_{\pm}) = S_{\pm}, \quad S (V_- (x)) = S_x, \quad S (V_- (x) \setminus \{x \}) = S_x^* $$
 for all~$x \in V_+$, and similar notation for the cost.
 Also, the conditional probability of an event~$A$ given the distribution of the origin~$\Or$ are written as
 $$ \begin{array}{rcl}
        P_x (A) & \n = \n & P (A \,| \,\Or = x) \vspace*{4pt} \\
      P_x^- (A) & \n = \n & P (A \,| \,\Or = \uniform (V_- (x) \setminus \{x \})) \vspace*{4pt} \\
        P^+ (A) & \n = \n & P (A \,| \,\Or = \uniform (V_+ \setminus \{0 \})) \vspace*{4pt} \\
        P^- (A) & \n = \n & P (A \,| \,\Or = \uniform (V_- \setminus V_- (0))) \end{array} $$
 and similar notation for the conditional mean, variance and covariance.
 The superscript~$+$ emphasizes that the contagion starts from a smart contract, while the superscript~$-$ emphasizes that the contagion starts from a user.
 The mean and variance of the loss resulting from a contagion in scenario~$j$ are written~$\mu_j$ and~$\sigma^2_j$, respectively.
 In particular,
 $$ \mu_1 = E_0 (C), \quad \mu_2 = E_0^- (C - C (\{\Or \})), \quad \mu_3 = E^+ (C_0), \quad \mu_4 = E^- (C_0) $$
 and similarly for the variance.
 We first recall a result from~\cite{jevtic_lanchier_2018} that will be useful later to study the mean and variance in scenarios~1 and~2.
 In the absence of users, i.e.,
 $$ X_- = 0 \quad \hbox{and} \quad G = \T_R = (V_+, E_+), $$
 in which case the parameter~$q$ and the distribution~$\widehat C_-$ are unimportant, the mean and the variance of the size of the contagion have already
 been studied in detail by the authors in~\cite{jevtic_lanchier_2018}.
 Using the notation above, their result gives exact expressions of the mean and the variance of the size~$S_+$ given that the contagion starts from
 the root of the graph.
 To state this result, let
 $$ \mu_+ = E (X_+) = \sum_{k = 0}^{\infty} \,k p_k \quad \hbox{and} \quad \sigma_+^2 = \var (X_+) = \sum_{k = 0}^{\infty} \,(k - \mu_+)^2 p_k $$
 be the mean and the variance of the number of edges starting from each smart contract and connecting two smart contracts.
 Comparing the size of the contagion with the number of individuals up to generation~$R$ in a certain branching process gives the following
 result from~\cite{jevtic_lanchier_2018}.
\begin{theorem} --
\label{th:branching}
 For a contagion starting at the root,
 $$ \begin{array}{rcl}
       E_0 (S_+) & \n = \n & \displaystyle \frac{1 - (\mu_+ p)^{R + 1}}{1 - \mu_+ p} \vspace*{8pt} \\
    \var_0 (S_+) & \n = \n & \displaystyle \frac{p (1 - p) \mu_+ + p^2 {\sigma_+^2}}{(1 - \mu_+ p)^2} \bigg(\frac{1 - (\mu_+ p)^{2R + 1}}{1 - \mu_+ p} - (2R + 1) (\mu_+ p)^R \bigg). \end{array} $$
\end{theorem}
 Our first result below shows how the conditional mean and variance of the total loss~$C$ relate to the conditional mean and variance of the size~$S_+$
 given that the contagion starts from a smart contract~$x \in V_+$.
 This can be used in combination with Theorem~\ref{th:branching} to obtain the conditional mean and variance in scenario~1.
 To state our result, as we did for~$X_+$, we let
 $$ \mu_- = E (X_-) = \sum_{k = 0}^{\infty} \,k q_k \quad \hbox{and} \quad \sigma_-^2 = \var (X_-) = \sum_{k = 0}^{\infty} \,(k - \mu_-)^2 q_k $$ 
 be the mean and the variance of the number of users connected to a given smart contract.
 By conditioning on the size~$S_+$, one can express the conditional mean and variance of the loss as a function of the conditional mean and
 variance of~$S_+$ as follows.
\begin{theorem} --
\label{th:connection}
 For all~$x \in V_+$,
 $$ \begin{array}{rcl}
       E_x (C) & \n = \n & E_x (S_+) E_0 (C_0) \vspace*{4pt} \\
    \var_x (C) & \n = \n & E_x (S_+) \var_0 (C_0) + \var_x (S_+) (E_0 (C_0))^2 \end{array} $$
 where the mean and variance of~$C_0$ are
 $$ \begin{array}{rcl}
       E_0 (C_0) & \n = \n & E (\widehat C_+) + q \mu_- \,E (\widehat C_-) \vspace*{4pt} \\
    \var_0 (C_0) & \n = \n & \var (\widehat C_+) + ({\sigma_-^2} - \mu_-)(q E (\widehat C_-))^2 + q \mu_- E ((\widehat C_-)^2). \end{array} $$
\end{theorem}
 Taking~$x = 0$ in the theorem gives the mean and variance of the total loss as a function of the mean and variance of~$S_+$ which, in turn, are given
 in Theorem~\ref{th:branching}.
 In particular, combining Theorems~\ref{th:branching} and~\ref{th:connection} directly gives the mean and variance in the first scenario.
 Even though the result is an obvious corollary of the first two theorems, we state it as a theorem for completeness.
\begin{theorem}[scenario 1] --
\label{th:case1}
 For all~$X_+$ and~$X_-$,
 $$ \begin{array}{rcl}
         \mu_1 & \n = \n & E_0 (S_+) E_0 (C_0) \vspace*{4pt} \\
    \sigma_1^2 & \n = \n & E_0 (S_+) \var_0 (C_0) + \var_0 (S_+) (E_0 (C_0))^2 \end{array} $$
 where the mean and variance of~$S_+$ and~$C_0$ are given in Theorems~\ref{th:branching} and~\ref{th:connection}.
\end{theorem}
 Recall that, in scenario~2, the contagion starts from one of the users of the root contract chosen uniformly at random.
 This user tries to compromise the network and the relevant loss consists of the cumulative cost of all the {compromised} vertices except for the originator.
 The key idea to study this scenario is to condition on the state of edge~$(0, x)$, where~$x$ is a user of the root contract, in order to express
 the conditional mean and variance of the loss given that the contagion starts from~$x$ as a function of the mean and variance given that the
 contagion starts from the root.
 Because the latter is known from Theorem~\ref{th:case1}, this leads to an explicit expression for the mean and variance in scenario~2.
 More precisely, we have the following theorem.
\begin{theorem}[scenario 2] --
\label{th:case2}
 For all~$X_+$ and~$X_-$,
 $$ \begin{array}{rcl}
         \mu_2 & \n = \n & q \,(\mu_1 - q E (\widehat C_-)) \vspace*{4pt} \\
    \sigma_2^2 & \n = \n & q \sigma_1^2 + q (1 - q)((\mu_1 - q E (\widehat C_-))^2 - qE (\widehat C_-))^2) \end{array} $$
 where~$\mu_1$ and~$\sigma_1^2$ are given in Theorem~\ref{th:case1}.
\end{theorem}
 In the last two scenarios, the objective is to study the loss~$C_0$ restricted to the root contract and its users when the contagion starts from
 outside this set.
 This is more difficult than the first two scenarios, but we can derive exact expressions in the context of deterministic graphs.
 Note however that the contagion process is still stochastic.
 More precisely, we assume that
 $$ P (X_+ = d_+) = P (X_- = d_-) = 1 \quad \hbox{for some} \quad d_+, d_- \in \N^*. $$
 In this case, $\mu_- = d_-$ and~$\sigma_-^2 = 0$, therefore Theorem~\ref{th:connection} gives
\begin{equation}
\label{eq:deterministic}
  \begin{array}{rcl}
     E_0 (C_0) & \n = \n & E (\widehat C_+) + q d_- \,E (\widehat C_-) \vspace*{4pt} \\
  \var_0 (C_0) & \n = \n & \var (\widehat C_+) - d_- (q E (\widehat C_-))^2 + q d_- E ((\widehat C_-)^2). \end{array}
\end{equation}
 Because in scenarios~3 and~4 the loss~$C_0$ is strictly positive if and only if the root gets {compromised}, it can be proved that the mean and
 variance when the contagion starts from outside~$V_- (0)$ are connected to the mean and variance in~\eqref{eq:deterministic} through the
 probability of the event
 $$ A = \hbox{the root is {compromised}} = \{0 \in \C (\Or) \}. $$
 The probability of this event and how the mean and variance in scenarios~3 and~4 are related to~\eqref{eq:deterministic} above are given
 in the following theorem.
\begin{theorem}[scenarios 3 and 4] --
\label{th:case34}
 For~$X_+ = d_+$ and~$X_- = d_-$,
 $$ P^+ (A) = p \,\bigg(\frac{1 - (d_+ p)^R}{1 - d_+ p} \bigg) \bigg(\frac{1 - d_+}{1 - d_+^R} \bigg) \quad \hbox{and} \quad P^- (A) = q P^+ (A). $$
 In scenarios~3 and~4, the mean and variance are then given by
 $$ \begin{array}{rclrcl}
    \mu_3 & \n = \n & P^+ (A) \,E_0 (C_0), & \quad \sigma_3^2 & \n = \n & P^+ (A) [P^+ (A^c) (E_0 (C_0))^2 + \var_0 (C_0)] \vspace*{4pt} \\
    \mu_4 & \n = \n & P^- (A) \,E_0 (C_0), & \quad \sigma_4^2 & \n = \n & P^- (A) [P^- (A^c) (E_0 (C_0))^2 + \var_0 (C_0)] \end{array} $$
 where~$E_0 (C_0)$ and~$\var_0 (C_0)$ are given in~\eqref{eq:deterministic}.
\end{theorem}
 Our last task is to combining Theorems~\ref{th:case1}--\ref{th:case34} to deduce the mean and variance of the aggregate financial loss up to time~$t$.
 Recall that scenario~$j$ occurs with probability~$Q_j$ at each arrival time of the Poisson process~$(N_t)$ independently of everything else, and let
 $$ \begin{array}{rcl}
      N_t^j & \n = \n & \hbox{number of occurrences of a type~$j$ contagion by time~$t$} \vspace*{2pt} \\
      L_t^j & \n = \n & \hbox{aggregate financial loss due to type~$j$ contagions by time~$t$}. \end{array} $$
 By the thinning property of Poisson processes, the processes~$(N_t^j)$ are independent Poisson processes with intensity~$\lambda Q_j$, from which
 it follows that
 $$ E (N_t^j) = \var (N_t^j) = \lambda t Q_j \quad \hbox{for} \quad 1 \leq j \leq 4. $$
 In particular, conditioning on~$N_t^j$, we get
\begin{equation}
\label{eq:aggregate-mean}
  E (L_t) = \sum_{j = 1}^4 \,E (L_t^j) = \sum_{j = 1}^4 \,E (E (L_t^j \,| \,N_t^j)) = \sum_{j = 1}^4 \lambda t Q_j \,\mu_j.
\end{equation}
 Using also that the contagions at different times (and therefore the loss resulting from these contagions) are independent, and applying the
 law of total variance,
\begin{equation}
\label{eq:aggregate-var}
  \var (L_t) = \sum_{j = 1}^4 \,(E (\var (L_t^j \,| \,N_t^j)) + \var (E (L_t^j \,| \,N_t^j))) = \sum_{j = 1}^4 \lambda t Q_j (\sigma_j^2 + \mu_j^2).
\end{equation}
 Combining~\eqref{eq:aggregate-mean} and~\eqref{eq:aggregate-var} with all our theorems gives explicit expressions for the
 mean and the variance of the random variable~$L_t$, as desired for the purpose of insurance pricing. \bigskip

\noindent {\bf Numerical results.}
\label{sec:numerics} {Under various parameter settings, we investigate the expectation~$E[L_t]$ and variance~${\var[L_t]}$ of loss distribution\footnote{{The characterization of~$E[L_t]$ and~${\var[L_t]}$ allow for straight forward calculation of actuarial fair risk, expectation principle based as well as standard deviation principle based risk premium for smart contract risk (\cite{embrechts2000actuarial}
 and~\cite{kaas2008modern}).}} of smart contract risk given the developed model. }\smallskip
%we perform pricing~$(P_t)$ of smart contract {(self-) insurance} under the assumptions of the developed model. Due to limitation of space, we present only premiums calculated using standard deviation principle.
 %Thus, we facilitate calculation of the premiums shown below:
 %$$ \begin{array}{rl}
 %   \hbox{actuarial fair premium:} & P_t = E [L_t] \vspace*{2pt} \\
 %   \hbox{expectation principle:} & P_t =  E [L_t] (1 + \delta) \vspace*{2pt} \\
 %   \hbox{standard deviation principle:} & P_t = E [L_t] + \delta \sqrt{\var [L_t]}. \end{array} $$
 
Without loss of generality we assume~$t = 1$ and have the profit loading factor~$\delta = 0.1$. Further, we assume~$\lambda = 1$, thus the attacks occur at rate one per unit of time\footnote{{In practice, for parameter~$\lambda$, a platform provider would use it's internal  statistics related to attack rates.}}. For illustrative purposes, two choices of probability mass function~$[p_0, p_1, p_2]$ for random smart contract tree edge formation are considered.
First, the smart contract tree with probabilistic formation of edges under consideration is characterized by the probability mass function~$[0, 0.4, 0.6]$.
 Hence, for each vertex, i.e., smart contract, the probability of zero offspring~$p_0$ with this choice is equal to zero.
 The probability of one offspring of a smart contract is~$p_1 = 0.4$ and the probability of two offspring of a smart contract is~$p_2 = 0.6$.
 Thus, a given smart contract will, with probability~0.4, have one offspring smart contract it communicates with and, with probability~0.6,
 two offspring smart contracts it communicates with.
 Second, we consider a deterministic smart contract tree characterized by the probability mass function of edge formation~{$[0, 0, 1]$}.
 In number of smart contract vertices, this tree dominates stochastically the above chosen tree with probabilistic formation,
 given the same radius~$R$.
 Across all experiments the common radius of the trees is chosen to be~$R = 3$. \\
%  Also, given the same radius, the chosen probabilistic tree, in terms of number of vertices, dominates the deterministic trees characterized by the
%  probability mass function of edge formation~{$[0, 1, 0, 0]$}.
\indent Under the assumption of log-normal distributions for both~$\widehat C_+$ and~$\widehat C_{-}$, we allow for three cost
 topologies (see Table~\ref{tab:costTable}).
 These costs materialize when smart contract and user wallets are compromised.
 The cost topologies under consideration are characterized by three cases of means and standard deviations.
{The choice of expectation of cost for the smart contracts (see second column in the table)
%  $E (\widehat C_+)$
 is stylized, kept to~10,000 monetary units, and made consistent across all cost cases.
 Similarly, the choice of expectation of cost for the users (see fourth column)
%  $E (\widehat C_-)$
 is stylized, kept to~1,000 monetary units, and made consistent across all cost cases as well.
 The standard deviation of cost for the smart contracts (see third column)
%  $\sqrt{\var (\widehat C_+)}$
 in a stylized fashion is allowed to change across cases, alternating between~0 and~5,000.
 Similarly, the standard deviation of cost for the users (see fifth column)
%  $\sqrt{\var (\widehat C_-)}$ in a stylized fashion
 is also allowed to change across cases, alternating between~0 and~500.}
 
 \begin{table}[h!]
{%\scriptsize
%\fontsize
\begin{tabular}{| c | c |  c  |  c  |  c  |  c  |  c | c |  c |  c |  }
\cline{2-5}
\multicolumn{1}{c|}{} & \multicolumn{2}{c|}{\bf Smart Contracts } & \multicolumn{2}{c|}{\bf Users } \\ \cline{2-5}
\multicolumn{1}{c|}{} &  Expectation   & Deviation & Expectation & Deviation \\ \cline{1-1}
\multicolumn{1}{|c|}{\bf Cost}  & of Cost & of Cost & of Cost & of Cost \\ \hline
\multirow{1}{*}{} & & & & \\  
\multirow{1}{*}{Topology} 	&	\multirow{1}{*}{$E[\widehat C_+]$}	&	\multirow{1}{*}{$\sqrt{\var[\widehat C_+]}$}	&	\multirow{1}{*}{$E[\widehat C_-]$}	&	\multirow{1}{*}{$\sqrt{\var[\widehat C_-]}$}	\\
& & & & 
\\
\hline
\multirow{1}{*}{I}	&10000	&0	&1000	&0	\\ \cdashline {1-5}[0.4pt/1pt]
\multirow{1}{*}{{II}}	&10000	&5000 	&1000	&0	\\ \cdashline {1-5}[0.4pt/1pt]
\multirow{1}{*}{III}	&10000	&0	&1000	&500	\\ \hline
\end{tabular}
\label{tab:pricesScenario1}
\caption{{\footnotesize Three cases of cost topology are given assuming the log-normal distribution for both $\widehat C_+$ and $\widehat C_-$.}}
\label{tab:costTable}}
\end{table}
 
\indent Two cases for the probability of smart contract edge contagion\footnote{{In practice, to choose the probability of the edge contagion~$p$, a platform provider (e.g. Ethereum) may perform risk classification by clustering their ecosystem of smart contracts across a predetermined set of features.
 The academic literature~\cite{grishchenko2018semantic, kalra2018zeus, tikhomirov2018smartcheck} or best
 practices (e.g. https://consensys.github.io/smart-contract-best-practices/) can guide the choice of such features.
 Alternatively, to create best practices, platform providers should consult smart contract audit providers that perform pre deployment smart contract analysis
 and consulting.
 Regretfully, the true value of $p$ is unknowable and practically unattainable.
 Thus, for found risk classes, according to their riskiness, and based on its judgment, a platform provider should impute values of edge contagion given their own internal expert knowledge.}} are considered:
 low probability of edge contagion characterized by~$p = 0.2$ and high probability of edge contagion characterized by~$p = 0.8$.
 Likewise, two cases for the probability of user edge contagion\footnote{{{In practice, to choose the probability of the edge contagion $q$, a platform provider should make considerations similar to when choosing parameter $p$.}}} are considered:
 low probability of user edge contagion characterized by~$q = 0.2$ and high probability of user edge contagion characterized by~$q = 0.8$. \\
\indent Within this numerical experiment setting, to calculate the moments for loss distribution of smart contract risk, we perform ten million simulations.
 Our unreported trials confirm that this number of simulations is sufficient to achieve desired prices stability and accuracy.
{This is additionally supported by the congruency between the simulation based results and the analytical results, as shown in Tables~\ref{tab:priceTable1}
 and~\ref{tab:priceTable2} where the difference between simulation and analytical results does not exceed one percent.
 For the sake of brevity, and because our findings based on numerical simulations are the same in all four scenarios, we only investigate the first
 and third scenarios}\footnote{The simulation results for scenario~2 and scenario~4 are available on request.}. \\
%  We investigate scenario~1 and scenario~3.
%  For the sake of brevity, scenario~2 and scenario~4 are omitted
%  Our unreported experiments concerning these experiments show that this decision does not result in a considerable loss of intuition presented
\indent Recall that, in scenario~1 (see Table~\ref{tab:priceTable1}), the contagion starts from the root contract.
 Here, there are several insights that can be deduced from our numerical results.
\begin{itemize}
\item {First, everything else being fixed, offspring distributions that result in a stochastically higher number of vertices (for both smart contracts
       and users) consistently lead to higher means and variances.}
       Hence, the stochastic ``size'' of the {interactions} impacts the moments: the bigger the network, the higher the moments.
%      {This can be proved rigorously using a so-called coupling argument, but our numerical results also show that the standard deviation principle follows this rule as well, which is more difficult to prove.} 
\smallskip
% \item First, for fixed probabilities of edge contagion, for pairs of probabilities defining smart contract tree and user star graph which, result in
%       stochastically higher number of vertices (both of smart contracts and users), the actuarial fair premium is consistently higher.
%       Hence, the stochastic ``size'' of the network impacts the pricing: the bigger the network, the higher the premium.
%       The standard deviation principle follows this rule as well. \smallskip
\item {Second, across all parameter settings, the fact that
      both smart contracts and users costs change in time (which is captured by random variables with differently parameterized distributions) makes
     {an} impact on the moments of loss distribution.
      Further, as expected, increasing the variance of the the costs while keeping their expectation fixed results in an increase of the variance of loss distribution.
      Also, the impact of variability of costs of smart contracts {versus} variability of costs of users is, in principle, different.}
%      In general, {the} higher the variability of the costs, the higher the .}
\end{itemize}
 Recall that, in scenario~3 (see Table~\ref{tab:priceTable2}), the origin of the contagion is chosen uniformly at random from the set of smart contracts
 other than the root.
{Following the analytical results, the simulations for this scenario were only performed when the tree-stars graph is deterministic.}
 Here, several insights can be deduced from our numerical results.
\begin{itemize}
\item {First, across all parameter settings in scenario~3, when compared to the corresponding settings in scenario~1, we observe significantly
      lower moments of the loss distribution.}
     %{This justifies the idea that this scenario represents a  of its own.}
\item {Second, similarly to scenario~1, higher the levels of contagion parameters higher the moments.}
\end{itemize}
Given the {high dimension of the} parameter space of the model and the number of scenarios, many more numerical investigations are conceivable.
 They are not given here both because of the constrains of space and because of the essentially intuitively obvious impact of the parameters.
 More importantly, we {point out} that the analytical results can be used to obtain the exact values of the moments in the context of
 scenarios~1 and~2, whereas they are limited to deterministic smart contracts/users networks in the context of scenarios~3 and~4.
 In contrast, the simulation based approach {does not suffer any such constraints}.
{In addition, the almost perfect match between our analytical and numerical results in the cases covered by our theorems validates our numerical
 results.
 In particular, our simulations are reliable enough to give extremely good approximations of the moments in all four scenarios and for any possible choices of the random tree-stars graph and network topology.} \smallskip

\begin{sidewaystable}[h!]
% \begin{table}[h!]
% \begin{adjustwidth}{-2.2cm}{}
{
%\scriptsize
\centering
\tiny
\begin{tabular}{| c | c | c | c |  c | c  |  c  |  c  |  }
\cline{5-8}
\multicolumn{2}{c}{} & \multicolumn{1}{c}{}&\multicolumn{1}{c|}{} & \multicolumn{2}{c|}{\bf Analytical Results} & \multicolumn{2}{c|}{\bf Simulation Results} \\ \cline{5-8}
\multicolumn{2}{c}{} & \multicolumn{1}{c}{}& & {}      &                   & {}      &          \\ \cline{1-4}
\multicolumn{2}{|l}{{Number of simulations: 10000000.}} & \multicolumn{1}{c}{}& & {\bf Expectation}   & {\bf Deviation} & {\bf Expectation}   & {\bf Deviation} \\ \cline{1-4}
\multicolumn{4}{|l|}{Contagion scenario: $[Q_1,Q_2,Q_3,Q_4] = [1.0,0.0,0.0,0.0]$}  & {\bf of Loss} & {\bf of Loss}    & {\bf of Loss} & {\bf of Loss}   \\ \hline
\multirow{1}{*}{\bf Cost} & {\bf Smart contracts} & {\bf Users} & {\bf Contagion} & & &  &  \\  \cline{1-4}
\multirow{2}{*}{Topology} & \multirow{2}{*}{$[\;p_0, \;p_1, \;p_2]$}	& \multirow{2}{*}{$[\;q_0, \;q_1, \;q_2, \;q_3\;, \;q_4\;]$}	& \multirow{2}{*}{$(p,q)$}	&	\multirow{1}{*}{$E[L_t]$}	&	\multirow{1}{*}{$\sqrt{\var[L_t]}$} &	\multirow{1}{*}{$E[L_t]$}	&	\multirow{1}{*}{$\sqrt{\var[L_t]}$}	\\
& & & &  & & &\\
\hline
\multirow{1}{*}{}	&[0.0,0.0,1.0]	&[0.0,0.0,0.0,0.0,1.0]	&	(0.8,0.8)	&68112.00	&21666.32	&68103.05	&21668.17	\\
\multirow{1}{*}{}	&[0.0,0.0,1.0]	&[0.0,0.1,0.2,0.3,0.4]	&	(0.8,0.8)	&63984.00	&20423.47	&63986.51	&20429.11	\\
\multirow{1}{*}{}	&[0.0,0.4,0.6]	&[0.0,0.0,0.0,0.0,1.0]	&	(0.8,0.8)	&51722.88	&21560.68	&51729.05		&21564.09\\
\multirow{1}{*}{}	&[0.0,0.4,0.6]	&[0.0,0.1,0.2,0.3,0.4]	&	(0.8,0.8)	&48588.16	&20307.61		&48589.85	&20304.28\\  \cdashline {2-8}[0.4pt/1pt]
\multirow{1}{*}{}	&[0.0,0.0,1.0]	&[0.0,0.0,0.0,0.0,1.0]	&	(0.8,0.2)	&55728.00	&17757.75		&55726.09	&17764.73\\
\multirow{1}{*}{}	&[0.0,0.0,1.0]	&[0.0,0.1,0.2,0.3,0.4]	&	(0.8,0.2)	&54696.00	&17414.61		&54695.96	&17412.82	\\
\multirow{1}{*}{}	&[0.0,0.4,0.6]	&[0.0,0.0,0.0,0.0,1.0]	&	(0.8,0.2)	&42318.72	&17664.04	&42308.19	&17666.34\\
\multirow{1}{*}{}	&[0.0,0.4,0.6]	&[0.0,0.1,0.2,0.3,0.4]	&	(0.8,0.2)	&41535.04	&17326.01		&41532.69	&17326.06\\ \cdashline {2-8}[0.4pt/1pt]
\multirow{1}{*}{{\large I}}	&[0.0,0.0,1.0]	&[0.0,0.0,0.0,0.0,1.0]	&	(0.2,0.8)	&20592.00	&11514.53		&20590.19	&11514.84	\\
\multirow{1}{*}{}	&[0.0,0.0,1.0]	&[0.0,0.1,0.2,0.3,0.4]	&	(0.2,0.8)	&19344.00	&10856.65	&19340.23	&10853.55\\
\multirow{1}{*}{}	&[0.0,0.4,0.6]	&[0.0,0.0,0.0,0.0,1.0]	&	(0.2,0.8)	&18775.68	&9816.03		&18779.33		&9820.13\\
\multirow{1}{*}{}	&[0.0,0.4,0.6]	&[0.0,0.1,0.2,0.3,0.4]	&	(0.2,0.8)	&17637.76		&9263.84		&17638.57		&9260.89\\ \cdashline {2-8}[0.4pt/1pt]
\multirow{1}{*}{}	&[0.0,0.0,1.0]	&[0.0,0.0,0.0,0.0,1.0]	&	(0.2,0.2)	&16848.00	&9438.48		&16848.34	&9437.55\\
\multirow{1}{*}{}	&[0.0,0.0,1.0]	&[0.0,0.1,0.2,0.3,0.4]	&	(0.2,0.2)	&16536.00	&9255.56		&16535.89	&9257.05\\
\multirow{1}{*}{}	&[0.0,0.4,0.6]	&[0.0,0.0,0.0,0.0,1.0]	&	(0.2,0.2)	&15361.92		&8050.01		&15365.43	&8055.28\\
\multirow{1}{*}{}	&[0.0,0.4,0.6]	&[0.0,0.1,0.2,0.3,0.4]	&	(0.2,0.2)	&15077.44		&7892.24		&15077.99		&7893.62	\\ \hline
\multirow{1}{*}{}	&[0.0,0.0,1.0]	&[0.0,0.0,0.0,0.0,1.0]	&	(0.8,0.8)	&68112.00	&24462.81	&68099.27	&24461.01	\\
\multirow{1}{*}{}	&[0.0,0.0,1.0]	&[0.0,0.1,0.2,0.3,0.4]	&	(0.8,0.8)	&63984.00	&23369.17	&63964.57	&23373.77\\
\multirow{1}{*}{}	&[0.0,0.4,0.6]	&[0.0,0.0,0.0,0.0,1.0]	&	(0.8,0.8)	&51722.88	&23723.89	&51731.94		&23722.85\\
\multirow{1}{*}{}	&[0.0,0.4,0.6]	&[0.0,0.1,0.2,0.3,0.4]	&	(0.8,0.8)	&48588.16	&22591.12		&48592.92	&22584.33\\ \cdashline {2-8}[0.4pt/1pt]
\multirow{1}{*}{}	&[0.0,0.0,1.0]	&[0.0,0.0,0.0,0.0,1.0]	&	(0.8,0.2)	&55728.00	&21079.32		&55729.30	&21082.61\\
\multirow{1}{*}{}	&[0.0,0.0,1.0]	&[0.0,0.1,0.2,0.3,0.4]	&	(0.8,0.2)	&54696.00	&20791.07 	&54694.60	&20789.54\\
\multirow{1}{*}{}	&[0.0,0.4,0.6]	&[0.0,0.0,0.0,0.0,1.0]	&	(0.8,0.2)	&42318.72	&20247.92	&42319.76	&20247.80\\
\multirow{1}{*}{}	&[0.0,0.4,0.6]	&[0.0,0.1,0.2,0.3,0.4]	&	(0.8,0.2)	&41535.04	&19953.72	&41533.41		&19953.31\\ \cdashline {2-8}[0.4pt/1pt]
\multirow{1}{*}{{\large II}}	&[0.0,0.0,1.0]	&[0.0,0.0,0.0,0.0,1.0]	&	(0.2,0.8)	&20592.00	&13099.02	&20588.87	&13099.83\\
\multirow{1}{*}{}	&[0.0,0.0,1.0]	&[0.0,0.1,0.2,0.3,0.4]	&	(0.2,0.8)	&19344.00	&12524.65	&19340.46	&12525.41	\\
\multirow{1}{*}{}	&[0.0,0.4,0.6]	&[0.0,0.0,0.0,0.0,1.0]	&	(0.2,0.8)	&18775.68	&11485.40	&18769.41		&11478.62	\\
\multirow{1}{*}{}	&[0.0,0.4,0.6]	&[0.0,0.1,0.2,0.3,0.4]	&	(0.2,0.8)	&17637.76		&11017.20		&17640.30	&11023.83	\\ \cdashline {2-8}[0.4pt/1pt]
\multirow{1}{*}{}	&[0.0,0.0,1.0]	&[0.0,0.0,0.0,0.0,1.0]	&	(0.2,0.2)	&16848.00	&11317.46		&16848.22	&11314.42	\\
\multirow{1}{*}{}	&[0.0,0.0,1.0]	&[0.0,0.1,0.2,0.3,0.4]	&	(0.2,0.2)	&16536.00	&11165.37		&16537.00	&11164.44	\\
\multirow{1}{*}{}	&[0.0,0.4,0.6]	&[0.0,0.0,0.0,0.0,1.0]	&	(0.2,0.2)	&15361.92		&10018.12		&15366.09	&10021.09\\
\multirow{1}{*}{}	&[0.0,0.4,0.6]	&[0.0,0.1,0.2,0.3,0.4]	&	(0.2,0.2)	&15077.44		&9891.79		&15078.96	&9894.34	\\ \hline
\multirow{1}{*}{}	&[0.0,0.0,1.0]	&[0.0,0.0,0.0,0.0,1.0]	&	(0.8,0.8)	&68112.00	&21666.32	&68112.36	&21757.41	\\
\multirow{1}{*}{}	&[0.0,0.0,1.0]	&[0.0,0.1,0.2,0.3,0.4]	&	(0.8,0.8)	&63984.00	&20423.47	&63989.54	&20492.10\\
\multirow{1}{*}{}	&[0.0,0.4,0.6]	&[0.0,0.0,0.0,0.0,1.0]	&	(0.8,0.8)	&51722.88	&21560.68	&51716.16		&21634.02\\
\multirow{1}{*}{}	&[0.0,0.4,0.6]	&[0.0,0.1,0.2,0.3,0.4]	&	(0.8,0.8)	&48588.16	&20307.61		&48579.16	&20368.48\\ \cdashline {2-8}[0.4pt/1pt]
\multirow{1}{*}{}	&[0.0,0.0,1.0]	&[0.0,0.0,0.0,0.0,1.0]	&	(0.8,0.2)	&55728.00	&17757.75		&55731.34	&17786.87	\\
\multirow{1}{*}{}	&[0.0,0.0,1.0]	&[0.0,0.1,0.2,0.3,0.4]	&	(0.8,0.2)	&54696.00	&17414.61		&54692.31	&17438.72	\\
\multirow{1}{*}{}	&[0.0,0.4,0.6]	&[0.0,0.0,0.0,0.0,1.0]	&	(0.8,0.2)	&42318.72	&17664.04	&42317.99	&17686.78\\
\multirow{1}{*}{}	&[0.0,0.4,0.6]	&[0.0,0.1,0.2,0.3,0.4]	&	(0.8,0.2)	&41535.04	&17326.01		&41537.19		&17338.98\\ \cdashline {2-8}[0.4pt/1pt]
\multirow{1}{*}{{\large III}}	&[0.0,0.0,1.0]	&[0.0,0.0,0.0,0.0,1.0]	&	(0.2,0.8)	&20592.00	&11514.53		&20591.05	&11567.02	\\
\multirow{1}{*}{}	&[0.0,0.0,1.0]	&[0.0,0.1,0.2,0.3,0.4]	&	(0.2,0.8)	&19344.00	&10856.65	&19345.33	&10903.39\\
\multirow{1}{*}{}	&[0.0,0.4,0.6]	&[0.0,0.0,0.0,0.0,1.0]	&	(0.2,0.8)	&18775.68	&9816.03		&18774.70		&9875.13\\
\multirow{1}{*}{}	&[0.0,0.4,0.6]	&[0.0,0.1,0.2,0.3,0.4]	&	(0.2,0.8)	&17637.76		&9263.84		&17636.26	&9308.82	\\ \cdashline {2-8}[0.4pt/1pt]
\multirow{1}{*}{}	&[0.0,0.0,1.0]	&[0.0,0.0,0.0,0.0,1.0]	&	(0.2,0.2)	&16848.00	&9438.48		&16844.12	&9452.76	\\
\multirow{1}{*}{}	&[0.0,0.0,1.0]	&[0.0,0.1,0.2,0.3,0.4]	&	(0.2,0.2)	&16536.00	&9255.56		&16533.49	&9265.07	\\
\multirow{1}{*}{}	&[0.0,0.4,0.6]	&[0.0,0.0,0.0,0.0,1.0]	&	(0.2,0.2)	&15361.92		&8050.01		&15356.57	&8062.96	\\
\multirow{1}{*}{}	&[0.0,0.4,0.6]	&[0.0,0.1,0.2,0.3,0.4]	&	(0.2,0.2)	&15077.44		&7892.24		&15076.54	&7904.62 \\ \hline
\end{tabular}
\caption{{\footnotesize {The analytically calculated and simulation based  first and second moments of loss distribution for smart contract risk.} The contagion scenario under consideration is scenario 1. The cost topologies I, II, and III are investigated. Simulation based results are achieved with 10 million simulation scenarios. The $t=1$ and $\lambda =1$ are assumed.}}
\label{tab:priceTable1}}
% \end{adjustwidth}
% \end{table}
\end{sidewaystable}

\begin{sidewaystable}[h!]
% \begin{table}[h!]
% \begin{adjustwidth}{-2.2cm}{}
{
%\scriptsize
\centering
\tiny
\begin{tabular}{| c | c | c | c |  c | c  |  c  |  c  |  c| }
\cline{5-9}
\multicolumn{2}{c}{} & \multicolumn{1}{c}{}&\multicolumn{1}{c|}{} & \multicolumn{2}{c|}{\bf Analytically Calculated Premium } & \multicolumn{2}{c|}{\bf Simulation Based Premium } \\ \cline{5-9}
%\multicolumn{2}{c}{} & \multicolumn{1}{c}{}& & {\bf Actuarial} &                   &                 \\
\multicolumn{2}{c}{} & \multicolumn{1}{c}{}& & {}      &                    & {}      &                    \\ \cline{1-4}
\multicolumn{2}{|l}{{Number of simulations: 10000000.}} & \multicolumn{1}{c}{}& & {\bf Expectation}   & {\bf Deviation} & {\bf Expectation}   & {\bf Deviation}  \\ \cline{1-4}
\multicolumn{4}{|l|}{Contagion scenario: $[Q_1,Q_2,Q_3,Q_4] = [1.0,0.0,0.0,0.0]$}  & {\bf of Loss} & {\bf of Loss}   & {\bf of Loss} & {\bf of Loss}   \\ \hline
\multirow{1}{*}{\bf Cost} & {\bf Smart contracts} & {\bf Users} & {\bf Contagion} & & &  & \\  \cline{1-4}
\multirow{2}{*}{Topology} & \multirow{2}{*}{$[\;p_0, \;p_1, \;p_2]$}	& \multirow{2}{*}{$[\;q_0, \;q_1, \;q_2, \;q_3\;, \;q_4\;]$}	& \multirow{2}{*}{$(p,q)$}	&	\multirow{1}{*}{$E[L_t]$}	&	\multirow{1}{*}{$\sqrt{\var[L_t]}$}	&	\multirow{1}{*}{$E[L_t]$}	&	\multirow{1}{*}{$\sqrt{\var[L_t]}$}\\
& & & & & &  & 
\\
\hline
\multirow{1}{*}{}			&[0.0,0.0,1.0]	&[0.0,0.0,0.0,0.0,1.0]	&	(0.8,0.8)	&9152.00	&6122.99	&9151.82	&6122.94	\\
\multirow{1}{*}{{\large I}}		&[0.0,0.0,1.0]	&[0.0,0.0,0.0,0.0,1.0]	&	(0.8,0.2)	&7488.00	&5024.34	&7487.80	&5023.78	\\
\multirow{1}{*}{}			&[0.0,0.0,1.0]	&[0.0,0.0,0.0,0.0,1.0]	&	(0.2,0.8)	&1232.00	&3847.64	&1231.56	&3847.14	\\
\multirow{1}{*}{}			&[0.0,0.0,1.0]	&[0.0,0.0,0.0,0.0,1.0]	&	(0.2,0.2)	&1008.00	&3151.20	&1008.40	&3151.87	\\ \hline
\multirow{1}{*}{}			&[0.0,0.0,1.0]	&[0.0,0.0,0.0,0.0,1.0]	&	(0.8,0.8)	&9152.00	&7404.35	&9151.30	&7404.43	\\
\multirow{1}{*}{{\large II}}		&[0.0,0.0,1.0]	&[0.0,0.0,0.0,0.0,1.0]	&	(0.8,0.2)	&7488.00	&6525.13	&7486.59	&6524.63	\\
\multirow{1}{*}{}			&[0.0,0.0,1.0]	&[0.0,0.0,0.0,0.0,1.0]	&	(0.2,0.8)	&1232.00	&4139.76	&1229.37	&4133.33	\\
\multirow{1}{*}{}			&[0.0,0.0,1.0]	&[0.0,0.0,0.0,0.0,1.0]	&	(0.2,0.2)	&1008.00	&3501.91	&1007.76	&3500.73	\\ \hline
\multirow{1}{*}{}			&[0.0,0.0,1.0]	&[0.0,0.0,0.0,0.0,1.0]	&	(0.8,0.8)	&9152.00	&6168.12	&9152.88	&6168.27	\\
\multirow{1}{*}{{\large III}}		&[0.0,0.0,1.0]	&[0.0,0.0,0.0,0.0,1.0]	&	(0.8,0.2)	&7488.00	&5038.12	&7490.12	&5036.94	\\
\multirow{1}{*}{}			&[0.0,0.0,1.0]	&[0.0,0.0,0.0,0.0,1.0]	&	(0.2,0.8)	&1232.00	&3857.33	&1232.50	&3857.91	\\
\multirow{1}{*}{}			&[0.0,0.0,1.0]	&[0.0,0.0,0.0,0.0,1.0]	&	(0.2,0.2)	&1008.00	&3154.16	&1007.68	&3153.85	\\ \hline
\end{tabular}
\caption{{\footnotesize {The analytically calculated and simulation based first and second moments of loss distribution for smart contract risk.} The contagion scenario under consideration is scenario 3. The cost topologies I, II, and III are investigated. Simulation based results are achieved with 10 million simulation scenarios. Only deterministic graph structure is considered. The $t=1$ and $\lambda =1$ are assumed.}}
\label{tab:priceTable2}}
% \end{adjustwidth}
% \end{table}
\end{sidewaystable}

\noindent {\bf Conclusion.} {In this paper, we develop a dynamic structural
 percolation model for the aggregate loss distribution due to cyber attacks on and contagious failures of smart contracts assuming
 a tree-stars topology of smart contracts and their users.
By focusing on network topologies where cycles are not allowed and by imposing, based on percolation theory, parsimonious contagion
 processes on such networks, coupled with the introduction of a topology of costs, we distinguish four different use cases or scenarios.
 Based on them, we robustly reduce the complexity of smart contract risk phenomena and allow for its effective modeling and loss distribution characterization. From a modeling standpoint, we allow for the dynamic nature of smart contracts and their users' topology, as well as temporal uncertainty of costs both for smart contracts and users, which is captured using random variables with various distributions.
 Within a rigorous mathematical framework through probabilistic analysis, we characterize the mean and variance, which are the main aspects of the loss distribution of smart contract risk. Because smart contract risk may represent a significant emerging liability for platform providers, companies and individuals which adopt this technology, our work can prove to be of considerable value to decision-makers while simultaneously supporting the penetration of this nascent technology in the economy, and thus unleashing its new potentials. }
 
\indent There are two immediate opportunities for further research following this work. First, modeling smart contract risk in a general star-fully connected graph, to account for loops in call graphs. {Second, modeling a collection (or ecosystem) of smart contracts with random interconnections, in order to ultimately characterize the aggregate risk smart contract platform providers can face.}

\section{Proofs}

\subsection{Theorem~\ref{th:connection}.}
 In this subsection, we prove Theorem~\ref{th:connection} which shows how the mean and variance of the total loss relate to the mean and variance
 of the number~$S_+$ of smart contracts being compromised.
 In this section and the next ones, we will repeatedly use that
 $$ S (A) = \sum_{y \in A} \,\zeta (y) \quad \hbox{and} \quad C (A) = \sum_{y \in A} \,\widehat C_y \,\zeta (y) \quad \hbox{for all} \quad A \subset V $$
 where~$\zeta : V \to \{0, 1 \}$ is the function
 $$ \zeta (y) = \ind \{\hbox{vertex~$y$ is {compromised}} \} = \ind \{y \in \C (\Or) \}. $$
 First, we compute the conditional mean and variance of the loss restricted to a smart contract and its users, given that the contagion starts
 from this smart contract, which corresponds to the second set of equations in the theorem.
\begin{lemma} --
\label{lem:connection-star}
 For all~$x \in V_+$,
 $$ \begin{array}{rcl}
       E_x (C_x) & \n = \n & E_0 (C_0) = E (\widehat C_+) + q \mu_- \,E (\widehat C_-) \vspace*{4pt} \\
    \var_x (C_x) & \n = \n & \var_0 (C_0) = \var (\widehat C_+) + ({\sigma_-^2} - \mu_-)(q E (\widehat C_-))^2 + q \mu_- E ((\widehat C_-)^2). \end{array} $$
\end{lemma}
\begin{proof}
 To begin with, we write
\begin{equation}
\label{eq:connection-star-1}
  C_x = C (V_- (x)) = \sum_{y \in V_- (x)} \widehat C_y \,\zeta (y) = \widehat C_x \,\zeta (x) + \sum_{y \in V_- (x) \setminus \{x \}} \widehat C_y \,\zeta (y).
\end{equation}
 Note also that, given that the contagion starts at smart contract~$x \in V_+$, each of the users of this contract, say~$y$, is {compromised}
 with probability~$q$, therefore
\begin{equation}
\label{eq:connection-star-2}
  \zeta (y) = \bernoulli (p) \quad \hbox{whenever} \quad \Or = x \in V_+ \ \hbox{and} \ y \in V_- (x) \setminus \{x \}.
\end{equation}
 Using~\eqref{eq:connection-star-1} and~\eqref{eq:connection-star-2}, and conditioning on~$Z_x = \card (V_- (x) \setminus \{x \})$,
\begin{equation}
\label{eq:connection-star-3}
  E_x (C_x \,| \,Z_x) = E_x (\widehat C_x \,\zeta (x)) + Z_x E_x (\widehat C_y \,\zeta (y)) = E (\widehat C_+) + Z_x \,q E (\widehat C_-)
\end{equation}
 while using also independence,
\begin{equation}
\label{eq:connection-star-4}
  \begin{array}{rcl}
  \var_x (C_x \,| \,Z_x) & \n = \n & \var_x (\widehat C_x \,\zeta (x)) + Z_x \var_x (\widehat C_y \,\zeta (y)) \vspace*{4pt} \\
                         & \n = \n & \var (\widehat C_x) + Z_x \,\big[E (\widehat C_y^2) \,E_x (\zeta (y)^2) - (E (\widehat C_y) \,E_x (\zeta (y)))^2 \big] \vspace*{4pt} \\
                         & \n = \n & \var (\widehat C_+) + Z_x \,\big[q E ((\widehat C_-)^2) - (q E (\widehat C_-))^2 \big]. \end{array}
\end{equation}
 Taking the expected value in~\eqref{eq:connection-star-3} gives
 $$ \begin{array}{rcl}
      E_x (C_x) & \n = \n & E (E_x (C_x \,| \,Z_x)) = E (E (\widehat C_+) + q Z_x \,E (\widehat C_-)) \vspace*{4pt} \\
                & \n = \n & E (\widehat C_+) + q E (X_-) \,E (\widehat C_-) = E (\widehat C_+) + q \mu_- \,E (\widehat C_-), \end{array} $$
 which proves the first part of the lemma.
 Using the law of total variance and adding the expected value of~\eqref{eq:connection-star-4} and the variance of~\eqref{eq:connection-star-3},
 we also get
 $$ \begin{array}{rcl}
    \var_x (C_x) & \n = \n & E (\var_x (C_x \,| \,Z_x)) + \var (E_x (C_x \,| \,Z_x)) \vspace*{4pt} \\
                 & \n = \n & \var (\widehat C_+) + E (X_-) \,\big[q E ((\widehat C_-)^2) - (q E (\widehat C_-))^2 \big] + \var (X_-) (q E (\widehat C_-))^2 \vspace*{4pt} \\
                 & \n = \n & \var (\widehat C_+) + \mu_- \,\big[q E ((\widehat C_-)^2) - (q E (\widehat C_-))^2 \big] + {\sigma_-^2} \,(q E (\widehat C_-))^2 \vspace*{4pt} \\
                 & \n = \n & \var (\widehat C_+) + ({\sigma_-^2} - \mu_-)(q E (\widehat C_-))^2 + q \mu_- E ((\widehat C_-)^2), \end{array} $$
 which proves the second part of the lemma.
\end{proof} \\ \\
 We now show how the mean and variance of the total loss across the network relate to the mean and variance of the number of {compromised} smart
 contracts, and the mean and variance in Lemma~\ref{lem:connection-star}, which corresponds to the first set of equations in the theorem.
\begin{lemma} --
\label{lem:connection-cost}
 For all~$x \in V_+$,
 $$ \begin{array}{rcl}
       E_x (C) & \n = \n & E_x (S_+) E_0 (C_0) \vspace*{4pt} \\
    \var_x (C) & \n = \n & E_x (S_+) \var_0 (C_0) + \var_x (S_+) (E_0 (C_0))^2. \end{array} $$
\end{lemma}
\begin{proof}
 Due to the independence of the state (open or closed) of the edges, and the independence of the local costs attached to the vertices, we have
\begin{equation}
\label{eq:connection-cost}
  E_x (C \,| \,S_+) = S_+ E_0 (C_0) \quad \hbox{and} \quad \var_x (C \,| \,S_+) = S_+ \var_0 (C_0)
\end{equation}
 The first equation in~\eqref{eq:connection-cost} implies that
 $$ E_x (C) = E_x (E_x (C \,| \,S_+)) = E_x (S_+) E_0 (C_0) $$
 while using also the second equation in~\eqref{eq:connection-cost} and the law of total variance,
 $$ \begin{array}{rcl}
    \var_x (C) & \n = \n & E_x (\var_x (C \,| \,S_+)) + \var_x (E_x (C \,| \,S_+)) \vspace*{4pt} \\
               & \n = \n & E_x (S_+ \var_0 (C_0)) + \var_x (S_+ E_0 (C_0)) \vspace*{4pt} \\
               & \n = \n & E_x (S_+) \var_0 (C_0) + \var_x (S_+) (E_0 (C_0))^2. \end{array} $$
 This completes the proof.
\end{proof} \\ \\
 Theorem~\ref{th:connection} is a direct consequence of Lemmas~\ref{lem:connection-star} and~\ref{lem:connection-cost}.

\subsection{Theorem~\ref{th:case2} (scenario~2).}
\label{sec:case2}
 This subsection deals with scenario~2 where the contagion starts from one of the users of the root chosen uniformly at random.
 This user compromises the system and the relevant loss is the cumulative cost of all the compromised vertices, except for the originator.
 To begin with, we prove the theorem when the contagion starts from a deterministic vertex~$x$ who is a user of the smart contract
 at the root.
 The main idea is to condition on whether
 $$ (0, x) = \hbox{edge connecting the root~0 and user~$x \in V_- (0) \setminus \{0 \}$} $$
 is open or closed in order to derive a relationship between the mean and variance of the loss when the contagion starts from~$x$ and their
 counterparts when the contagion starts from the root, for which an explicit expression is known from Theorem~\ref{th:case1}.
\begin{lemma} --
\label{lem:case2-mean0}
 For all~$x \in V_- (0) \setminus \{0 \}$,
 $$ E_0 (C - C (\{x \})) = \mu_1 - q E (\widehat C_-). $$
\end{lemma}
\begin{proof}
 Because the loss~$C (\{x \}) = 0$ whenever edge~{$e = (0, x)$} is closed and the contagion starts at the root, and that edge~$e$ is open with probability~$q$,
 $$ \begin{array}{rcl}
      E_0 (C - C (\{x \})) & \n = \n & E_0 (C) - E_0 (C (\{x \}) \,| \,\xi (e) = 1) \,P_0 (\xi (e) = 1) \vspace*{4pt} \\
                           & \n = \n & E_0 (C) - q E (\widehat C_-) = \mu_1 - q E (\widehat C_-). \end{array} $$
 This completes the proof.
\end{proof} \\ \\
 We now prove a weak version of the first part of Theorem~\ref{th:case2} with the contagion starting from a specific user of the root rather
 than a user chosen uniformly at random.
\begin{lemma} --
\label{lem:case2-mean}
 For all~$x \in V_- (0) \setminus \{0 \}$,
 $$ E_x (C - C (\{x \})) = q (E_0 (C) - q E (\widehat C_-)). $$
\end{lemma}
\begin{proof}
 Because~$C - C (\{x \}) = 0$ when~{$e = (0, x)$} is closed,
\begin{equation}
\label{eq:case2-mean}
  \begin{array}{rcl}
    E_x (C - C (\{x \}) \,| \,\xi (e)) & \n = \n & E_x (C - C \{x \} \,| \,\xi (e) = 1) \,\ind \{\xi (e) = 1 \} \vspace*{4pt} \\
                                       & \n = \n & E_0 (C - C (\{x \})) \,\xi (e). \end{array}
\end{equation}
 Taking the expected value and applying Lemma~\ref{lem:case2-mean0}, we conclude
 $$ \begin{array}{rcl}
      E_x (C - C (\{x \})) & \n = \n &
      E_x (E_x (C - C (\{x \}) \,| \,\xi (e))) =
      E_0 (C - C (\{x \})) \,E_x (\xi (e)) \vspace*{4pt} \\ & \n = \n &
      q E_0 (C - C (\{x \})) =
      q (E_0 (C) - q E (\widehat C_-)) =
      q (\mu_1 - q E (\widehat C_-)). \end{array} $$
 This completes the proof.
\end{proof} \\ \\
 We now study the {variance} of the loss.
\begin{lemma} --
\label{lem:case2-var0}
 For all~$x \in V_- (0) \setminus \{0 \}$,
 $$ \var_0 (C - C (\{x \})) = \sigma_1^2 - q (1 - q)(E (\widehat C_-))^2. $$
\end{lemma}
\begin{proof}
 Let~$y \in V$, $y \neq x$.
 Because the unique self-avoiding path connecting~$x$ and~$y$ goes through the root, given that the contagion starts from the root, the events
 that~$x$ gets {compromised} and that~$y$ gets {compromised} are independent.
 This implies that
 $$ \cov_0 (\zeta (x), \zeta (y)) = 0 \quad \hbox{for all} \quad y \neq x. $$
 Since in addition the~$\widehat C_z$ are independent,
 $$ \cov_0 (C, C (\{x \})) = \sum_{y \in V} \,\cov_0 (\widehat C_y \,\zeta (y), \widehat C_x \,\zeta (x)) = \var_0 (\widehat C_x \,\zeta (x)). $$
 Using also that~$\zeta (x) = \bernoulli (q)$ when the contagion starts at the root,
 $$ \begin{array}{rcl}
    \var_0 (C - C (\{x \})) & \n = \n & \var_0 (C) + \var_0 (C (\{x \})) - 2 \cov_0 (C, C (\{x \})) \vspace*{4pt} \\
                            & \n = \n & \sigma_1^2 + \var_0 (\widehat C_x \,\zeta (x)) - 2 \var_0 (\widehat C_x \,\zeta (x)) \vspace*{4pt} \\
                            & \n = \n & \sigma_1^2 - q (1 - q)(E (\widehat C_x))^2. \end{array} $$
 Recalling that~$\widehat C_x = \widehat C_-$ in distribution, the result follows.
\end{proof}
\begin{lemma} --
\label{lem:case2-var}
 For all~$x \in V_- (0) \setminus \{0 \}$,
 $$ \var_x (C - C (\{x \})) = q \sigma_1^2 + q (1 - q)((\mu_1 - q E (\widehat C_-))^2 - qE (\widehat C_-))^2). $$
\end{lemma}
\begin{proof}
 As in the proof of Lemma~\ref{lem:case2-mean}, {letting~$e = (0, x)$},
\begin{equation}
\label{eq:case2-var}
  \begin{array}{rcl}
  \var_x (C - C (\{x \}) \,| \,\xi (e)) & \n = \n & \var_x (C - C \{x \} \,| \,\xi (e) = 1) \,\ind \{\xi (e) = 1 \} \vspace*{4pt} \\
                                        & \n = \n & \var_0 (C - C (\{x \})) \,\xi (e). \end{array}
\end{equation}
 Using~\eqref{eq:case2-mean} and~\eqref{eq:case2-var}, and the law of total variance, we get
 $$ \begin{array}{rcl}
    \var_x (C - C (\{x \})) & \n = \n &
       E_x (\var_x (C - C (\{x \}) \,| \,\xi (e))) \vspace*{4pt} \\ & \n & \hspace*{40pt} + \ \var_x (E_x (C - C (\{x \}) \,| \,\xi (e))) \vspace*{4pt} \\ & \n = \n &
    \var_0 (C - C (\{x \})) \,E_x (\xi (e)) \vspace*{4pt} \\ & \n & \hspace*{40pt} + \ (E_0 (C - C (\{x \})))^2 \,\var_x (\xi (e)) \vspace*{4pt} \\ & \n = \n &
      q \var_0 (C - C (\{x \})) + q (1 - q)(E_0 (C - C (\{x \})))^2. \end{array} $$
 Then, applying Lemmas~\ref{lem:case2-mean0} and~\ref{lem:case2-var0}, we conclude that
 $$ \begin{array}{rcl}
    \var_x (C - C (\{x \})) & \n = \n &
      q (\sigma_1^2 - q (1 - q)(E (\widehat C_-))^2) + q (1 - q)(\mu_1 - q E (\widehat C_-))^2  \vspace*{4pt} \\ & \n = \n &
      q \sigma_1^2 + q (1 - q)((\mu_1 - q E (\widehat C_-))^2 - q E (\widehat C_-))^2) . \end{array} $$
 This completes the proof.
\end{proof} \\ \\
{ Because the expressions of the mean and variance given in Lemmas~\ref{lem:case2-mean} and~\ref{lem:case2-var} are constant across all possible
 choices of the user~$x$ of the smart contract at the root, the two lemmas hold more generally when the origin of the contagion is chosen
 uniformly at random from the set of all users of the root, a general result proved in the next lemma for any random variable.
\begin{lemma} --
\label{lem:uniform}
 Let~$X$ be any random variable such that
 $$ E_x (X) = \mu \quad \hbox{and} \quad \var_x (X) = \sigma^2 \quad \hbox{for all} \quad x \in V_- (0) \setminus \{0 \}. $$
 Then~$E_0^- (X) = \mu$ and~$\var_0^- (X) = \sigma^2$.
\end{lemma}
\begin{proof}
 Let~$G = (V, E)$ be a realization of the random graph, and let~$n$ be the number of users of the root for this realization.
 Observe that
\begin{equation}
\label{eq:uniform-1}
  \begin{array}{rcl}
     E (X \,| \,\Or) & \n = \n & \displaystyle \sum_x \,E (X \,| \,\Or = x) \,\ind \{\Or = x \} = \sum_x \,E_x (X) \,\ind \{\Or = x \} \vspace*{4pt} \\
  \var (X \,| \,\Or) & \n = \n & \displaystyle \sum_x \,\var (X \,| \,\Or = x) \,\ind \{\Or = x \} = \sum_x \,\var_x (X) \,\ind \{\Or = x \} \end{array}
\end{equation}
 where the sums are over the set~$V_- (0) \setminus \{0 \}$. Then,
 $$ \begin{array}{rcl}
      E_0^- (X) = E_0^- (E (X \,| \,\Or)) & \n = \n & \displaystyle \sum_x \,E_x (X) \,P_0^- (\Or = x) \vspace*{4pt} \\
                                          & \n = \n & \displaystyle \mu \,\sum_x \,P_0^- (\Or = x) = \mu. \end{array} $$
 Similarly, taking the mean in the second equation in~\eqref{eq:uniform-1} gives
\begin{equation}
\label{eq:uniform-2}
  \begin{array}{rcl}
    E_0^- (\var (X \,| \,\Or)) & \n = \n & \displaystyle \sum_x \,\var_x (X) \,P_0^- (\Or = x) \vspace*{4pt} \\
                               & \n = \n & \displaystyle \sigma^2 \,\sum_x \,P_0^- (\Or = x) = \sigma^2. \end{array}
\end{equation}
 Also, using that the covariance is
 $$ \begin{array}{l}
      \cov_0^- (E_x (X) \,\ind \{\Or = x \}, E_y (X) \,\ind \{\Or = y \}) \vspace*{4pt} \\ \hspace*{20pt} = \
      \mu^2 (P_0^- (\Or = x, \Or = y) - P_0^- (\Or = x) P_0^- (\Or = y)) \vspace*{4pt} \\ \hspace*{40pt} = \
      \mu^2 (P_0^- (\Or = x) \,\ind \{x = y \} - P_0^- (\Or = x) P_0^- (\Or = y)), \end{array} $$
 taking the variance in the first equation in~\eqref{eq:uniform-1} gives
\begin{equation}
\label{eq:uniform-3}
  \var_0^- (E (X \,| \,\Or)) = \sum_x \bigg(\frac{\mu^2}{n} \bigg) - \sum_{x, y} \bigg(\frac{\mu^2}{n^2} \bigg) = \mu^2 - \mu^2 = 0.
\end{equation}
 Using~\eqref{eq:uniform-2} and~\eqref{eq:uniform-3}, and the law of total variance, we get
 $$ \var_0^- (X) = E_0^- (\var (X \,| \,\Or)) + \var_0^- (E (X \,| \,\Or)) = \sigma^2. $$
 This completes the proof.
\end{proof} \\ \\
 The theorem directly follows from Lemmas~\ref{lem:case2-mean}, \ref{lem:case2-var}, and~\ref{lem:uniform}.}

%%%%%%%%%%%%%%%%%%%%%%%%%%%%%%%%%%%%%%%%%%%%%%%%%%%%%%%%%%%%%%%%%%%%%%%%%%%%%%%%%%%%%%%%%%%%%%%%%%%%%%%%%%%%%%%%%%%%%%%%%%%%%%%%%%%%%%%%%%%%%%%%%%%%%%%%%%%%%%%%%%%%%%%%%%%%

\subsection{Theorem~\ref{th:case34} (scenarios~3 and~4).}
\label{sec:case34}
 Recall that, in scenarios~3 and~4, the contagion starts from a vertex outside~$V_- (0)$, in which case the loss we are interested in is the
 cumulative loss~$C_0$ of the smart contract at the root and the users of this smart contract.
 The starting point and common idea behind the proof of all the statements in Theorem~\ref{th:case34} is the following:
 for all vertices
 $$ x \in V \setminus V_- (0) \quad \hbox{and} \quad y \in V_- (0), $$
 the unique self-avoiding path connecting~$x$ and~$y$ must go through the root.
 This implies that, when the contagion starts from vertex~$x$, the loss~$C_0 \neq 0$ only if the root gets {compromised}.
 In particular, the conditional expected loss in both scenarios reduces to
\begin{equation}
\label{eq:mean-34}
  \begin{array}{rcl}
    E^{\pm} (C_0 \,| \,\zeta (0)) & \n = \n & E^{\pm} (C_0 \,| \,\zeta (0) = 0)(1 - \zeta (0)) + E^{\pm} (C_0 \,| \,\zeta (0) = 1) \,\zeta (0) \vspace*{4pt} \\
                                  & \n = \n & E^{\pm} (C_0 \,| \,\zeta (0) = 1) \,\zeta (0) = E_0 (C_0) \,\zeta (0), \end{array}
\end{equation}
 and similarly for the conditional variance,
\begin{equation}
\label{eq:var-34}
  \var^{\pm} (C_0 \,| \,\zeta (0)) = \var^{\pm} (C_0 \,| \,\zeta (0) = 1) \,\zeta (0) = \var_0 (C_0) \,\zeta (0).
\end{equation}
 Equations~\eqref{eq:mean-34} and~\eqref{eq:var-34} indicate that the mean and variance of the loss can be expressed using the mean and variance of
 the random {variable}~$\zeta (0)$.
 We now focus on scenario~3 in which the contagion starts from a smart contract chosen uniformly at random among all the smart contracts excluding the root.
 To begin with, we compute the mean of~$\zeta (0)$, which is the probability of~$A$.
\begin{lemma} --
\label{lem:mean-zeta+}
 For~$X_+ = d_+$ and~$X_- = d_-$,
 $$ E^+ (\zeta (0)) = P^+ (A) = p \,\bigg(\frac{1 - (d_+ p)^R}{1 - d_+ p} \bigg) \bigg(\frac{1 - d_+}{1 - d_+^R} \bigg). $$
\end{lemma}
\begin{proof}
 To simplify the notation, we introduce
 $$ \psi (d_+, p) = \frac{\phi (d_+ p)}{\phi (d_+)} \quad \hbox{where} \quad \phi (a) = \sum_{r = 0}^{R - 1} \,a^r = \frac{1 - a^R}{1 - a} $$
 and denote by~$D = d (0, \Or)$ the distance between the root and the origin of the contagion.
 Using that the number of contracts at distance~$r$ from the root is~$d^r$, the probability mass function of the distance~$D$ can be written as follows:
 for all~$r = 1, 2, \ldots, R$,
\begin{equation}
\label{eq:mean-zeta+1}
  P^+ (D = r) = \frac{d_+^r}{\card (V_+ \setminus \{0 \})} = \frac{d_+^r}{d_+ + d_+^2 + \cdots + d_+^R} = \frac{d_+^{r - 1}}{\phi (d_+)}.
\end{equation}
 Note also that, given that the contagion starts at~$x \in V_+$, the root gets {compromised} if and only if the unique self-avoiding path from~$x$
 to the root is open.
 Because this path has~$d (0, x)$ edges and those edges are independently open with probability~$p$, we get
\begin{equation}
\label{eq:mean-zeta+2}
  P^+ (A \,| \,D = r) = P^+ (\zeta (0) = 1 \,| \,D = r) = p^r.
\end{equation}
 Combining~\eqref{eq:mean-zeta+1} and~\eqref{eq:mean-zeta+2}, we deduce that
 $$ P^+ (A) = \sum_{r = 1}^R \ \frac{d_+^{r - 1} p^r}{\phi (d_+)}
            = \frac{p}{\phi (d_+)} \ \sum_{r = 0}^{R - 1} \,d_+^r p^r
            = p \ \frac{\phi (d_+ p)}{\phi (d_+)}
            = p \,\psi (d_+, p). $$
 Recalling the definition of~$\phi$ and~$\psi$, the lemma follows.
\end{proof} \\ \\
 The next natural step is to compute the variance of~$\zeta (0)$.
 To do so, we will use the following preliminary result about the covariance.
\begin{lemma} --
\label{lem:cov-D}
 For all~$r, s = 1, 2, \ldots, R$,
 $$ \cov^+ (\ind \{D = r \}, \ind \{D = s \}) = \frac{d_+^{r - 1}}{\phi (d_+)} \bigg(\ind \{r = s \} - \frac{d_+^{s - 1}}{\phi (d_+)} \bigg). $$
\end{lemma}
\begin{proof}
 Observing that
 $$ \begin{array}{rcl}
    \cov^+ (\ind \{D = r \}, \ind \{D = s \}) & \n = \n & P^+ (D = r, D = s) - P^+ (D = r) P^+ (D = s) \vspace*{4pt} \\
                                              & \n = \n & P^+ (D = r) \,(\ind \{r = s \} - P^+ (D = s)) \end{array} $$
 and recalling from~\eqref{eq:mean-zeta+1} that~$P^+ (D = r) = d_+^{r - 1} / \phi (d_+)$ give the result.
\end{proof}
\begin{lemma} --
\label{lem:var-zeta+}
 For~$X_+ = d_+$ and~$X_- = d_-$,
 $$ \var^+ (\zeta (0)) = P^+ (A) P^+ (A^c) \quad \hbox{where} \quad P^+ (A) = p \,\bigg(\frac{1 - (d_+ p)^R}{1 - d_+ p} \bigg) \bigg(\frac{1 - d_+}{1 - d_+^R} \bigg). $$
\end{lemma}
\begin{proof}
 By~\eqref{eq:mean-zeta+2}, we have~$\zeta (0) = \bernoulli (p^r)$ when~$D = r$ therefore
\begin{equation}
\label{eq:var-zeta+1}
  E^+ (\zeta (0) \,| \,D = r) = p^r \quad \hbox{and} \quad \var^+ (\zeta (0) \,| \,D = r) = p^r (1 - p^r).
\end{equation}
 Using Lemma~\ref{lem:cov-D} and the first equation in~\eqref{eq:var-zeta+1}, we get
\begin{equation}
\label{eq:var-zeta+2}
  \begin{array}{l}
  \displaystyle \var^+ (E^+ (\zeta (0) \,| \,D)) =
  \displaystyle \var^+ \bigg(\sum_{r = 1}^R \,p^r \,\ind \{D = r \} \bigg) \vspace*{4pt} \\ \hspace*{25pt} =
  \displaystyle \sum_{r = 1}^R \,\frac{d_+^{r - 1} p^{2r}}{\phi (d_+)} - \sum_{r, s = 1}^R \,\frac{d_+^{r + s - 2} p^{r + s}}{(\phi (d_+))^2} =
  \displaystyle \sum_{r = 1}^R \,\frac{d_+^{r - 1} p^{2r}}{\phi (d_+)} - \bigg(\sum_{r = 1}^R \,\frac{d_+^{r - 1} p^r}{\phi (d_+)} \bigg)^2 \vspace*{6pt} \\ \hspace*{25pt} =
  \displaystyle p^2 \ \frac{\phi (d_+ p^2)}{\phi (d_+)} - p^2 \bigg(\frac{\phi (d_+ p)}{\phi (d_+)} \bigg)^2 =
  \displaystyle p^2 \,\psi (d_+, p^2) - (P^+ (A))^2. \end{array}
\end{equation}
 Using~\eqref{eq:mean-zeta+1} and the second equation in~\eqref{eq:var-zeta+1}, we get
\begin{equation}
\label{eq:var-zeta+3}
  \begin{array}{l}
  \displaystyle E^+ (\var^+ (\zeta (0) \,| \,D)) =
  \displaystyle \sum_{r = 1}^R \,p^r (1 - p^r) P^+ (D = r) \vspace*{0pt} \\ \hspace*{25pt} =
  \displaystyle \frac{1}{\phi (d_+)} \ \sum_{r = 1}^R \,p^r (1 - p^r) \,d_+^{r - 1} =
  \displaystyle \frac{1}{\phi (d_+)} \ \sum_{r = 0}^{R - 1} \big(p (d_+ p)^r - p^2 (d_+ p^2)^r \big) \vspace*{6pt} \\ \hspace*{25pt} =
  \displaystyle p \ \frac{\phi (d_+ p)}{\phi (d_+)} - p^2 \ \frac{\phi (d_+ p^2)}{\phi (d_+)} =
  \displaystyle P^+ (A) - p^2 \,\psi (d_+, p^2). \end{array}
\end{equation}
 Using the law of total variance and~\eqref{eq:var-zeta+2}--\eqref{eq:var-zeta+3}, we conclude that
 $$ \begin{array}{rcl}
    \var^+ (\zeta (0)) & \n = \n & E^+ (\var^+ (\zeta (0) \,| \,D)) + \var^+ (E^+ (\zeta (0) \,| \,D)) \vspace*{4pt} \\
                       & \n = \n & P^+ (A) - (P^+ (A))^2 = P^+ (A)(1 - P^+ (A)) = P^+ (A) P^+ (A^c). \end{array} $$
 This completes the proof
\end{proof} \\ \\
 Using the previous results, we can now study the loss in scenario~3.
\begin{lemma} --
\label{lem:mean-3}
 For~$X_+ = d_+$ and~$X_- = d_-$,
 $$ \begin{array}{rcl}
         \mu_3 & \n = \n & P^+ (A) \,E_0 (C_0) \vspace*{4pt} \\
    \sigma_3^2 & \n = \n & P^+ (A) [P^+ (A^c) (E_0 (C_0))^2 + \var_0 (C_0)]. \end{array} $$
\end{lemma}
\begin{proof}
 Taking the expected value in~\eqref{eq:mean-34}, we get
 $$ \begin{array}{rcl}
    \mu_3 = E^+ (C_0) & \n = \n & E^+ (E^+ (C_0 \,| \,\zeta (0))) \vspace*{4pt} \\
                      & \n = \n & E_0 (C_0) \,E^+ (\zeta (0)) = P^+ (A) \,E_0 (C_0). \end{array} $$
 To find the variance, we take the expected value in~\eqref{eq:var-34} to get
\begin{equation}
\label{eq:mean-3-1}
  E^+ (\var^+ (C_0 \,| \,\zeta (0))) = \var_0 (C_0) \,E^+ (\zeta (0)) = P^+ (A) \,\var_0 (C_0)
\end{equation}
 and take the variance in~\eqref{eq:mean-34} and apply Lemma~\ref{lem:var-zeta+} to get
\begin{equation}
\label{eq:mean-3-2}
  \begin{array}{rcl}
  \var^+ (E^+ (C_0 \,| \,\zeta (0))) & \n = \n & (E_0 (C_0))^2 \var^+ (\zeta (0)) \vspace*{4pt} \\
                                     & \n = \n & P^+ (A) P^+ (A^c) \,(E_0 (C_0))^2. \end{array}
\end{equation}
 Using the law of total variance and~\eqref{eq:mean-3-1}--\eqref{eq:mean-3-2}, we conclude that
 $$ \begin{array}{rcl}
    \sigma_3^2 & \n = \n & \var^+ (C_0) = E^+ (\var^+ (C_0 \,| \,\zeta (0))) + \var^+ (E^+ (C_0 \,| \,\zeta (0))) \vspace*{4pt} \\
               & \n = \n & P^+ (A) \,\var_0 (C_0) + P^+ (A) P^+ (A^c) \,(E_0 (C_0))^2 \vspace*{4pt} \\
               & \n = \n & P^+ (A) \big[P^+ (A^c) (E_0 (C_0))^2 + \var_0 (C_0) \big]. \end{array} $$
 This completes the proof.
\end{proof} \\ \\
 Finally, we deal with scenario~4 in which the contagion starts from a vertex chosen uniformly at random from the set of users excluding
 the users of the root contract.
 As previously, we start by computing the expected value of~$\zeta (0)$, which is the probability of~$A$.
\begin{lemma} --
\label{lem:mean-zeta-}
 For~$X_+ = d_+$ and~$X_- = d_-$,
 $$ E^- (\zeta (0)) = P^- (A) = q P^+ (A) = pq \,\bigg(\frac{1 - (d_+ p)^R}{1 - d_+ p} \bigg) \bigg(\frac{1 - d_+}{1 - d_+^R} \bigg). $$
\end{lemma}
\begin{proof}
 Because~$X_- = d_-$, all the contracts have the same number of users, from which it follows that the distance~$D$ in scenarios~3 and~4
 are related as follows:
\begin{equation}
\label{eq:mean-zeta-1}
  P^- (D = r + 1) = P^+ (D = r) \quad \hbox{for all} \quad r = 1, 2, \ldots, R.
\end{equation}
 In addition, for all~$x \in V_-$, the unique self-avoiding path from~$x$ to the root has~$d (0, x) - 1$ edges in the subset~$E_+$ and one
 edge in the subset~$E_-$ therefore
\begin{equation}
\label{eq:mean-zeta-2}
  P^- (A \,| \,D = r + 1) = p^r q = q P^+ (A \,| \,D = r).
\end{equation}
 Combining~\eqref{eq:mean-zeta-1} and~\eqref{eq:mean-zeta-2}, we conclude that
 $$ \begin{array}{l}
      P^- (A) = \displaystyle \sum_{r = 1}^R \,P^- (A \,| \,D = r + 1) P^- (D = r + 1) \vspace*{-4pt} \\ \hspace*{80pt} =
                \displaystyle \sum_{r = 1}^R \,q P^+ (A \,| \,D = r) P^+ (D = r) = q P^+ (A), \end{array} $$
 and the proof is complete.
\end{proof} \\ \\
 Using~\eqref{eq:mean-zeta-1}, and repeating the proof of Lemma~\ref{lem:cov-D}, give
\begin{equation}
\label{eq:case4-1}
  \cov^- (\ind \{D = r + 1 \}, \ind \{D = s + 1 \}) = \cov^+ (\ind \{D = r \}, \ind \{D = s \}).
\end{equation}
 It also follows from~\eqref{eq:mean-zeta-2} that~$\zeta (0) = \bernoulli (p^r q)$ when~$D = r + 1$ so
\begin{equation}
\label{eq:case4-2}
  \begin{array}{rcl}
     E^- (\zeta (0) \,| \,D = r + 1) & \n = \n & p^r q \vspace*{4pt} \\
  \var^- (\zeta (0) \,| \,D = r + 1) & \n = \n & p^r q (1 - p^r q). \end{array}
\end{equation}
 Repeating the proof of Lemma~\ref{lem:var-zeta+} using~\eqref{eq:case4-1}, and~\eqref{eq:case4-2} in place of~\eqref{eq:var-zeta+1},
\begin{equation}
\label{eq:case4-3}
  \begin{array}{rcl}
  \var^+ (\zeta (0)) & \n = \n & q P^+ (A) (1 - q P^+ (A)) \vspace*{4pt} \\
                     & \n = \n & P^- (A)(1 - P^- (A)) = P^- (A) P^- (A^c). \end{array}
\end{equation}
 Finally, using~\eqref{eq:mean-34} and~\eqref{eq:var-34} like in the proof of Lemma~\ref{lem:mean-3}, together with~\eqref{eq:case4-3}, we get the
 mean and variance given in the second part of Theorem~\ref{th:case34}.

%%%%%%%%%%%%%%%%%%%%%%%%%%%%%%%%%%%%%%%%%%%%%%%%%%%%%%%%%%%%%%%%%%%%%%%%%%%%%%%%%%%%%%%%%%%%%%%%%%%%%%%%%%%%%%%%%%%%%%%%%%%%%%%%%%%%%%%%%%%%%%%%%%%%%%%%%%%%%%%%%%%%%%%%%%%%

\section*{Acknowledgments}
{This work is partially supported by the NSF grant \#CNS-2000792.}
Additionally, Petar Jevti\'{c} expresses his gratitude to the ASU Center for Assured and Scalable Data Engineering, for their financial support and guidance. 

%%%%%%%%%%%%%%%%%%%%%%%%%%%%%%%%%%%%%%%%%%%%%%%%%%%%%%%%%%%%%%%%%%%%%%%%%%%%%%%%%%%%%%%%%%%%%%%%%%%%%%%%%%%%%%%%%%%%%%%%%%%%%%%%%%%%%%%%%%%%%%%%%%%%%%%%%%%%%%%%%%%%%%%%%%%%

\newpage
\bibliographystyle{plain}
\bibliography{references}

\begin{thebibliography}{10}

\bibitem{atzei2017survey}
Nicola Atzei, Massimo Bartoletti, and Tiziana Cimoli.
\newblock A survey of attacks on ethereum smart contracts (sok).
\newblock In {\em Principles of Security and Trust}, pages 164--186. Springer,
  2017.

\bibitem{broadbent_hammersley_1957}
S.~R. Broadbent and J.~M. Hammersley.
\newblock Percolation processes. {I}. {C}rystals and mazes.
\newblock {\em Proc. Cambridge Philos. Soc.}, 53:629--641, 1957.

\bibitem{chen2017under}
Ting Chen, Xiaoqi Li, Xiapu Luo, and Xiaosong Zhang.
\newblock Under-optimized smart contracts devour your money.
\newblock In {\em Software Analysis, Evolution and Reengineering (SANER), 2017
  IEEE 24th International Conference on}, pages 442--446. IEEE, 2017.

\bibitem{embrechts2000actuarial}
Paul Embrechts.
\newblock Actuarial versus financial pricing of insurance.
\newblock {\em The Journal of Risk Finance}, 1(4):17--26, 2000.

\bibitem{frowis2017code}
M~Frowis and R~Bohme.
\newblock In code we trust?: Measuring the control flow immutability of all
  smart contracts deployed on ethereum.
\newblock {\em LNCS}, 10436:357--372, 2017.

\bibitem{grimmett_1989}
G.~R. Grimmett.
\newblock {\em Percolation}.
\newblock Springer-Verlag, New York, 1989.

\bibitem{grishchenko2018semantic}
Ilya Grishchenko, Matteo Maffei, and Clara Schneidewind.
\newblock A semantic framework for the security analysis of ethereum smart
  contracts.
\newblock In {\em International Conference on Principles of Security and
  Trust}, pages 243--269. Springer, 2018.

\bibitem{iansiti2017truth}
Marco Iansiti and Karim~R Lakhani.
\newblock The truth about blockchain.
\newblock {\em Harvard Business Review}, 95(1):118--127, 2017.

\bibitem{insights2016banking}
CB~Insights.
\newblock Banking is only the start: 20 big industries where blockchain could
  be used.
\newblock {\em CB Insights}, 25, 2016.

\bibitem{jevtic_lanchier_2018}
Petar Jevti\'{c} and Nicolas Lanchier.
\newblock Dynamic structural percolation model of loss distribution for cyber
  risk of small and medium-sized enterprises for tree-based {LAN} topology.
\newblock {\em Insurance Math. Econom.}, 91:209--223, 2020.

\bibitem{kaas2008modern}
Rob Kaas, Marc Goovaerts, Jan Dhaene, and Michel Denuit.
\newblock {\em Modern actuarial risk theory: using R}, volume 128.
\newblock Springer Science \& Business Media, 2008.

\bibitem{kalra2018zeus}
Sukrit Kalra, Seep Goel, Mohan Dhawan, and Subodh Sharma.
\newblock Zeus: Analyzing safety of smart contracts.
\newblock NDSS, 2018.

\bibitem{lanchier2017stochastic}
Nicolas Lanchier.
\newblock {\em Stochastic modeling}.
\newblock Springer, 2017.

\bibitem{marcus2018low}
Yuval Marcus, Ethan Heilman, and Sharon Goldberg.
\newblock Low-resource eclipse attacks on ethereum's peer-to-peer network.
\newblock {\em IACR Cryptology ePrint Archive}, 2018:236, 2018.

\bibitem{mehar2019understanding}
Muhammad~Izhar Mehar, Charles~Louis Shier, Alana Giambattista, Elgar Gong,
  Gabrielle Fletcher, Ryan Sanayhie, Henry~M Kim, and Marek Laskowski.
\newblock Understanding a revolutionary and flawed grand experiment in
  blockchain: The dao attack.
\newblock {\em Journal of Cases on Information Technology (JCIT)},
  21(1):19--32, 2019.

\bibitem{nakamoto2008bitcoin}
Satoshi Nakamoto.
\newblock Bitcoin: A peer-to-peer electronic cash system.
\newblock 2008.

\bibitem{nikolic2018finding}
Ivica Nikoli{\'c}, Aashish Kolluri, Ilya Sergey, Prateek Saxena, and Aquinas
  Hobor.
\newblock Finding the greedy, prodigal, and suicidal contracts at scale.
\newblock In {\em Proceedings of the 34th Annual Computer Security Applications
  Conference}, pages 653--663, 2018.

\bibitem{ryder1979constructing}
Barbara~G Ryder.
\newblock Constructing the call graph of a program.
\newblock {\em IEEE Transactions on Software Engineering}, (3):216--226, 1979.

\bibitem{shevchenko2011modelling}
Pavel~V Shevchenko.
\newblock {\em Modelling operational risk using Bayesian inference}.
\newblock Springer Science \& Business Media, 2011.

\bibitem{shift2015technology}
Deep Shift.
\newblock Technology tipping points and societal impact.
\newblock In {\em World Economic Forum Survey Report}, 2015.

\bibitem{shrier2016blockchain}
David Shrier, Weige Wu, and Alex Pentland.
\newblock Blockchain \& infrastructure (identity, data security).
\newblock {\em Massachusetts Institute of Technology-Connection Science}, 1(3),
  2016.

\bibitem{swan2015blockchain}
Melanie Swan.
\newblock {\em Blockchain: Blueprint for a new economy}.
\newblock " O'Reilly Media, Inc.", 2015.

\bibitem{szabo1996smart}
Nick Szabo.
\newblock Smart contracts: building blocks for digital markets.
\newblock {\em EXTROPY: The Journal of Transhumanist Thought,(16)}, 1996.

\bibitem{tikhomirov2018smartcheck}
Sergei Tikhomirov, Ekaterina Voskresenskaya, Ivan Ivanitskiy, Ramil Takhaviev,
  Evgeny Marchenko, and Yaroslav Alexandrov.
\newblock Smartcheck: Static analysis of ethereum smart contracts.
\newblock 2018.

\bibitem{van2018blockchain}
Mark Van~Rijmenam and Philippa Ryan.
\newblock {\em Blockchain: Transforming Your Business and Our World}.
\newblock Routledge, 2018.

\end{thebibliography}

%\begin{thebibliography}{100}

%\bibitem{jevtic_lanchier_2018}
% Jevti\' c, P. and Lanchier, N. (2018).
% Dynamic structural percolation model of loss distribution for cyber risk of small and medium-sized enterprises for tree-based LAN topology. Preprint.

%\end{thebibliography}

%\begin{table}[h!]
%\begin{center}
%{
%\scriptsize
%\def\arraystretch{1.2}
%\begin{tabular}{| c | c  |  c  |  c  |  c  | }
%\cline{2-5}
%\multicolumn{1}{c|}{}   & \multicolumn{4}{c|}{$E_r(S)$ assuming $[\;p_0, \;p_1, \;p_2, \;p_3\;] = [0, 0, 0, 1]$} \\ \cline{2-5}
%\multicolumn{1}{c|}{}   & \multicolumn{2}{c|}{$p = 0.2$} & \multicolumn{2}{c|}{$p = 0.8$} \\ \cline{1-5}
%\multicolumn{1}{|c|}{$r$}  & { Analytical} 		& { Simulated}                & { Analytical}            	&{ Simulated} \\ \hline
%$ 0 $	& 2.177552	& 2.176000 & 22.976907 & 22.984000 \\	\hline
%$ 1 $	& 2.315189	& 2.316800 & 21.681858 & 21.684800 \\	\hline
%$ 2 $	& 2.001532	& 1.999360 & 18.561597 & 18.571840 \\	\hline
%$ 3 $	& 1.360507	& 1.359872 & 15.208322 & 15.217472 \\	\hline										
%\end{tabular}}
%% \label{tab:info}
%\end{center}
%\caption{{\footnotesize The analytical versus simulated values of the expected size of cyber attack due to breach $E_r(S)$ given level $r$ are presented. A deterministic tree ($d=3$) having four levels ($R=4$) was assumed. The two probabilities of edge contagion are considered and the simulation results are based on one million simulations.}}
%\label{tab:sizeTable}
%\end{table}

\end{document}